\documentclass[conference]{IEEEtran}
\usepackage{cite}
\usepackage[hyphens]{url}  
\usepackage[pdftex,bookmarksnumbered,hidelinks,breaklinks]{hyperref}

\usepackage{amsmath,amssymb,amsfonts}
\usepackage{algorithmic}
\usepackage{graphicx}
\usepackage{textcomp}
\usepackage{xcolor}
\usepackage{amsmath}
\usepackage{amssymb}
\usepackage{wrapfig,lipsum,booktabs}
\usepackage{diagbox}
\usepackage{etoolbox}
\usepackage{fancyhdr} 
\usepackage{caption}
\usepackage{tabularx}
\usepackage{comment}
\usepackage{changepage}
\usepackage{enumerate}
\usepackage[linesnumbered,ruled,vlined]{algorithm2e}
\usepackage[english]{babel}
\usepackage[utf8]{inputenc}

\usepackage{subfiles}

\newenvironment{SUBENVccomment}[2]{\color{#1}[#2:]~}{\color{black}}

\definecolor{author1}{rgb}     {0.9,0.5,0.0}
\definecolor{author2}{rgb}     {0.6,0.0,0.8}
\definecolor{author3}{rgb}     {0.0,0.5,0.0}
\definecolor{author4}{rgb}     {0.9,0.2,0.2}

\def\BibTeX{{\rm B\kern-.05em{\sc i\kern-.025em b}\kern-.08em
    T\kern-.1667em\lower.7ex\hbox{E}\kern-.125emX}}

\ifCLASSINFOpdf
\else
\fi

\begin{document}

\title{Practical Attacks on Voice Spoofing Countermeasures}

\author{\IEEEauthorblockN{Andre Kassis}
\IEEEauthorblockA{University of Waterloo\\
akassis@uwaterloo.ca}
\and
\IEEEauthorblockN{Urs Hengartner}
\IEEEauthorblockA{University of Waterloo\\
urs.hengartner@uwaterloo.ca}}

\IEEEoverridecommandlockouts
\makeatletter\def\@IEEEpubidpullup{6.5\baselineskip}\makeatother

\maketitle

\begin{abstract}
Voice authentication has become an integral part in security-critical operations, such as bank transactions and call center conversations. The vulnerability of automatic speaker verification systems (ASVs) to spoofing attacks instigated the development of countermeasures (CMs), whose task is to tell apart bonafide and spoofed speech. Together, ASVs and CMs form today's voice authentication platforms, advertised as an impregnable access control mechanism. We develop the first practical attack on CMs, and show how a malicious actor may efficiently craft audio samples to bypass voice authentication in its strictest form. Previous works have primarily focused on non-proactive attacks or adversarial strategies against ASVs that do not produce speech in the victim's voice. The repercussions of our attacks are far more severe, as the samples we generate sound like the victim, eliminating any chance of plausible deniability. Moreover, the few existing adversarial attacks against CMs mistakenly optimize spoofed speech in the feature space and do not take into account the existence of ASVs, resulting in inferior synthetic audio that fails in realistic settings. We eliminate these obstacles through our key technical contribution: a novel joint loss function that enables mounting advanced adversarial attacks against combined ASV/CM deployments directly in the time domain. Our adversarials achieve concerning black-box success rates against state-of-the-art authentication platforms (up to 93.57\%). Finally, we perform the first targeted, over-telephony-network attack on CMs, bypassing several challenges and enabling various potential threats, given the increased use of voice biometrics in call centers. Our results call into question the security of modern voice authentication systems in light of the real threat of attackers bypassing these measures to gain access to users' most valuable resources.
\end{abstract}


\hyphenation{ADVJOINT ADVCM ADVSR}

\section{Introduction}
Automatic speaker verification systems (ASVs) are widely used for authentication as platforms where a claimed identity is verified by comparing features extracted from a given audio sample against a ``voiceprint" obtained from  previously collected recordings. ASVs have gained popularity, primarily due to the convenience they offer, releasing users from the burden of having to remember passwords. 
The vulnerability of ASVs to spoofing attacks---impersonation~\cite{lau2004vulnerability}, replay~\cite{zhizhengwu}, speech synthesis~\cite{shen2018natural}, and voice conversion~\cite{todavoice, mukhopadhyay2015all}---has been the subject of numerous academic and industrial projects, which led to the development of spoofing countermeasures (CMs) through individual endeavors~\cite{ahmed2020void} or INTERSPEECH's ASVspoof challenge~\cite{wu2017asvspoof}. These works envision voice authentication platforms in the form of a  system combining an ASV and a CM. The former verifies an identity claim, while the latter rejects spoofed, machine-generated samples. Such combined solutions have become prevalent in deployed applications~\cite{das2020attacker}. Many service providers explicitly inform their customers of the existence of a CM alongside an ASV to mitigate security concerns associated with spoofing attacks~\cite{hsbc}.

Voice authentication has been deployed in security-critical environments, such as banks (e.g., Citi Bank~\cite{india} 
and first direct~\cite{firstdirect}). 
VocalPassword~\cite{vocalpassword} by Nuance Communications (acquired in April by Microsoft for \$19.7B) and similar products are widely used by service providers for access control, either in their call centers or to enable transactions using smartphone apps~\cite{nuance, nuancebank}. Boastful statements (e.g.,``No one else has a voice just like you''~\cite{td}) released by these service providers, promising their customers impermeable security raise questions regarding the soundness of these claims.


The main objective of this work is to investigate the robustness of CMs and their ability to turn ASVs into trustworthy authentication systems. The four spoofing attacks mentioned above against which CMs are robust are not proactive, but opportunistic in nature. Excluding impersonation, which does not pose realistic threats to ASVs~\cite{das2020attacker}, attacks belonging to the remaining three categories are obtained via benign algorithms and methods, not designed for defeating voice authentication systems, but typically developed for helping people with disabilities or for entertainment~\cite{serdiuk-2020}. The question is, hence, whether advanced attacks crafted with a malicious intent may compromise an ASV augmented with a CM.

CMs are mainly useful to protect against attacks wherein the adversary is stealthy and wants to sound like the victim to human listeners, which is an essential requirement when targeting security-critical systems (see \S\ref{sec:Threat}). While many works present proactive attacks on ASVs using adversarial examples~\cite{kreuk2018fooling, li2020adversarial, chen2019real, jati2021adversarial, du2020sirenattack} that may go undetected by a CM, these attacks do not achieve the stealthiness requirement (when examined by a human listener) and are not practical attacks against CM-equipped voice authentication platforms. 

This paper demonstrates the first practical attacks on CMs---an integral part of real-world voice authentication platforms. 
Our realistic threat model mandates that four separate components must be evaded by a successful attack: the CM, the underlying ASV, a text-to-speech unit examining the contents of the spoken phrase, and a human judge who may be asked to verify the machine's decision. We establish adversarial examples as a suitable attack strategy and formalize the problem of generating adversarials against CMs from spoofed speech as an optimization problem, constrained by the threat model.

We find that none of the existing works on attacks against CMs~\cite{liu2019adversarial, zhang2020black} 
manage to produce adversarial examples that succeed in practical settings. Our study 
attributes these failures to the adversarial perturbations crafted in the spectral domain not being able to withstand (inverse) time-frequency transformations, resulting in machine-like artefacts that CMs (specifically) can detect. Another factor existing works overlook is the existence of ASVs alongside the targeted CMs that need to be accounted for when generating an attack sample. 

We introduce a novel class of attacks against CMs by implementing the necessary logic for supporting the back propagation of gradients through feature extractors, enabling us to craft adversarial examples in the time domain---a requirement we show to be essential when attacking a CM. Our gradient-supported feature extractors enable us to robustify the attacks by taking the ASV into account through our joint loss function, which includes a designated part to regularize the adversarials and exclude perturbations that risk removing the victim's voiceprint. The optimization is performed on audio in the victim's voice, generated with off-the-shelf speech synthesis or voice conversion algorithms. The result is an end-to-end attack, bypassing voice authentication in its strictest form.


To demonstrate the practical risks of our attacks, we evaluate them in two real-world scenarios: the banking app attack and the phone call attack. In the former, the attacker bypasses the voice authentication system used by a banking app to guard in-app transactions~\cite{anz}. In the latter, the attacker poses as the victim in a phone call with the target system to issue false transactions~\cite{td}. The adversarials we present circumvent the obstacles that prevent adversarial attacks from succeeding over a phone call~\cite{abdullah2020faults} and defeat the effects of the telephony network distortion. Our attacks remain undetected even when inspected by a human, eliminating any chance of plausible deniability for the victim. As a result of our attacks and findings, users should rethink their trust in voice biometrics, and researchers and service providers should invest resources in developing methods that can make these mechanisms more secure. 

Our attacks are evaluated against CMs taken from the top submissions to the 2019 ASVspoof challenge~\cite{todisco2019asvspoof}. Our ASVs, deployed side-by-side with the CMs, consist of x-vector and i-vector GMM-UBM models, representing cutting-edge technology. Our large-scale study experiments with a variety of configurations as target systems and presents successful attacks in all of these settings. Our adversarials are all evaluated in the black-box scenario, exploiting the transferability of adversarial examples. We conduct a user study on MTurk \cite{amazon} to evaluate the ability of our optimized spoofed examples to fool human listeners. Our attacks achieve concerning success rates of up to 93.57\% in the black-box setting.

Overall, this work makes the following contributions:
\begin{itemize}
  \item We introduce and formalize a realistic threat model wherein spoofing countermeasures (CMs) are employed to secure voice authentication. We theoretically scrutinize the strategies proposed in the literature for crafting audio adversarial examples against CMs and find the existing approaches ineffective under the restrictions of the practical threat model we consider. We conduct experiments to corroborate these findings.
    \item We present the first technique for significantly degrading the performance of the most robust form of voice authentication through a novel class of adversarial attacks. 
    We develop a method to generate adversarial examples against CMs in the time domain. By minimizing a novel joint loss function, we generate powerful adversarials that evade four defenses simultaneously: CMs, ASVs, humans, and content verifiers. 
   \item We demonstrate the first \textit{\textbf{targeted}} over-telephony-network adversarial attack on combined ASV/CM systems and propose techniques to make it robust to network distortion.
   \item The end-to-end latency of our attacks is below 5 seconds, making them a practical real-time strategy for defeating voice authentication systems.
    \item We will make our source code publicly available for fellow researchers to reproduce our results.
\end{itemize}

\section{Background}
\label{sec:Background}
This section introduces the preliminaries of automatic speaker verification systems and spoofing countermeasures.

\subsection{Automatic Speaker Verification Systems}
At the very core of ASVs lies the voiceprint---a set of unique measurable features of an individual found in their voice. This voiceprint is believed to be an identifying characteristic since it captures physical factors such as the shape and size of the vocal tract and larynx, alongside a ``behavioral signature'' consisting of one's accent, rhythm, pronunciation and more~\cite{zheng2017robustness}. This purported property of voiceprints led researchers to investigate their validity as biometrics.

An ASV operates in two phases, namely an enrollment phase and a verification phase. During enrollment, the user supplies speech samples that are used to construct their voiceprint and derive the speaker's model, which will serve as their identifying signature onward. Upon future access requests, the user's identity is verified via provided speech samples that are checked against the voiceprint to output a decision (accept/reject). The verification stage can follow one of three schemes: 1) Text-dependent, where a predetermined phrase is repeatedly used for authentication, 2) Text-independent, where the user may provide any random phrase, or 3) Text-prompted, wherein the system requires a specific random text from the user to be spoken. The third option is the most secure due to its robustness to replay attacks, and is regarded as a robust defense strategy adopted by several service providers, making it the focus of this paper. Our attacks circumvent this strongest defense as well as all the other verification schemes.

Researchers have come up with various sets of features that have become the gold-standard for ASVs. These features include the Mel-frequency Cepstral Coefficients (MFCC) and Log Power Magnitude Spectrum (LPMS), with the former predominating the literature for its expressive power and ability to capture numerous speaker characteristics~\cite{li2020adversarial}. As for the architectures of ASVs, there are typically three representatives: i-vector speaker embedding-based systems, neural-networks (x-vector) based systems, and end-to-end approaches~\cite{li2020adversarial}. This work considers Gaussian Mixture Models systems based on i-vector and x-vector embeddings (or "GMM i-vector/x-vector")~\cite{li2020adversarial}\cite{du2020sirenattack} due to their popularity and wide applicability in biometric authentication tasks.

\subsection{Spoofing Countermeasures}
\label{sec:background:CMs}
ASV systems are vulnerable to spoofing attacks, where the voice samples used at the verification step are not provided by the genuine user whose identity is claimed, but rather an imposter. The difference between spoofing and zero-effort attacks is that in the latter the attacker merely claims to be the victim and provides a recording of their own voice as proof, hoping for a lucky confusion to occur. ASV systems are highly efficient in rejecting zero-effort attacks. 

Spoofing attacks can be classified into four categories: 1) Impersonation attacks, where the attacker manipulates their own voice in order to sound like the victim (which is accomplished without the help of a machine), 2) Voice synthesis attacks, where attackers use advanced ML technology~\cite{shen2018natural} to synthetically produce voice samples mimicking the victim, 3) Voice conversion attacks, where the attacker resorts to state-of-the-art technology to algorithmically convert their voice in a recording to that of the victim, while leaving the content intact, and 4) Replay attacks, where the attacker replays a previously recorded sample by the original user. In impersonation attacks, the attackers tend to make vocal caricatures of their target speakers by manipulating high-level audible cues more than the low-level spectral features used by ASVs. Hence, ASVs are naturally robust to impersonation attacks~\cite{das2020attacker}.

The need for reliable ASVs led to the emergence of spoofing countermeasures (CMs). These are classifiers for solving the binary problem of distinguishing between genuine speech samples (bonafide) and their spoofed counterparts. Similar to the task of speaker verification, spoofing countermeasures rely on a generalized set of characteristics that distinguish human from machine-generated speech, or artefacts that are only present when a computer is involved in the sample's generation. These nuances are captured by sets of features developed or studied for this purpose, including the MFCCs and LPMSs described earlier, in addition to the Constant-Q Cepstral Coefficients (CQCCs).

CMs are designed to be deployed with ASVs. A voice sample has to be approved by both systems to pass authentication.

\section{Threat Model}
\label{sec:Threat}

\textbf{Attacker's Goal:} The attacker wishes to impersonate some user of choice at a remote system. Interaction with the remote system may take place in two forms:
\begin{enumerate}
    \item Through a designated app - This setting has gained momentum over the past few years as many service providers (banks in particular) are offering "secure" voice-triggered services through their apps \cite{nuance}. The user issues a transaction and then authenticates it using their voice.
    \item Over telephony network - Similarly, there has been an increase in the number of banks and service providers offering remote services over the phone. The user calls the service provider and interacts with the system while their identity is verified using voice authentication~\cite{zoom}.
\end{enumerate}

Access to the remote system is protected with voice authentication. To mitigate the threats of spoofed speech, the system is also equipped with a CM~\cite{ahmed2020void}. We exclude text-dependent verification schemes since they are vulnerable to replay attacks. 
We assume the verification scheme to be overcome is either text-independent or text-prompted. In both cases, the attack becomes harder to mount since the spoken phrase must contain a very specific content; in the former case, the user may simply interact by saying something like: "Transfer 1,000\$ to account \#XXXX", which is used both for the purpose of verification and for conveying the intended command that must be preserved. In the latter case, the content presented by the remote system must be accurately repeated.

In the designated-app scenario, the recording provided by the user is forwarded directly to the remote server. In the over-telephony-network scenario, the input is encoded and then transmitted over the telephony network to the remote server. The second scenario introduces additional challenges, as the attacks will have to withstand the effects of the noisy medium.

Finally, and perhaps most importantly, the attacks must be stealthy. Since we only consider security-critical environments, it is natural to assume that suspicious transactions will be flagged and verified by a human agent before they are executed. Alternatively, if a malicious transaction gets executed, the victim will dispute the transaction, resulting in a retroactive investigation. Thus, the attacks must withstand an examination by a human judge. To that end, the attacker wants to sound like the victim to any human listener. 
\\

\textbf{Attacker's Knowledge:} \label{subsec:attacker_knowledge}
We are mainly concerned with the more realistic black-box setting, though we also demonstrate attacks in the white-box setting for reference. That is, the attacker does not have access to the target models' internals or to the training datasets used by the target systems (i.e., the speakers and utterances used to train those models). Recall that the target system employs CM models that require spoofed speech samples to train them. These spoofed samples are created with spoofing algorithms. We assume the attacker can use the same spoofing algorithms (but not the samples themselves) for the attack because the leading technology in this domain is the common standard in use. In Appendix~\ref{app:trained_on_eval} we show that this assumption does not result in an adversary more powerful than an attacker who does not know the spoofing algorithms that were part of the training of the target system.


Finally, the attacker is in possession of sufficient data to train speech synthesis models in the victim's voice. State-of-the-art speech synthesis systems accomplish this with only $\sim 15$ minutes of speech from the victim~\cite{tts}. Such data is easy to obtain thanks to social media, and especially for a determined attacker targeting a specific individual.
\\
    
\textbf{Attacker's Capabilities:} The attacker can bypass the microphone on the device used to interact with the remote server to inject adversarial audio waveforms directly into the service provider's app (or into the phone call app). This requires that the attacker obtains \textbf{\textit{some}} rooted device on which they can bypass the microphone, making this assumption highly plausible. Importantly, we do \textbf{not} assume the attacker has access to the victim's device. 


In the designated-app scenario, the attacker installs the victim's banking app on the rooted device. Most banks allow their users to install their apps on multiple devices. They may require the user to enter a password or a PIN~\cite{anz}, which the attacker can obtain through social engineering~\cite{datafloq}.  In the over-telephony-network scenario, this requirement can be waived, as the attacker may use any device to place the call without the need for knowing a PIN. Also, in this scenario there is no need for a rooted device as the attacker can use a computer and a VOIP program (e.g., Skype) to mount the attack. Although banking apps may be equipped with ``root detectors", these can often be evaded~\cite{security_2020}. It is worth noting that the attacker's ability to root the device does not make more powerful attacks possible, as the entire authentication process takes place on the server side~\cite{vocalpassword}.

The attacker has sufficient computational resources and data to build shadows that model the target system. The attacker trains those models and crafts attacks based on them.  

Overall, aside from computational resources, we only require the attacker to have access to a rooted device, 15 minutes of speech in the victim's voice, and perhaps a PIN obtained through social engineering, depending on the threat model.

\section{Attacking a Voice Authentication System}
\subsection{Problem Statement}

In our setting an attacker aims to produce an audio signal to bypass voice authentication. The threat model mandates that the attacker produce a signal capable of fooling both machines and human listeners. The machine consists of an ASV and a CM. The input to each of the two modules is a set of acoustic (spectral) features extracted from the signal. 

For an audio input $x \in X$, where $X$ is the space of audio signals, and a claimed identity $U_{ID}$, the ASV's decision is:
\begin{equation*}
    ASV(x,\; U_{ID}) = h_{ASV}(g_{ASV}(x), g_{ASV}(DB(U_{ID}))),
\end{equation*}

\noindent
where $h_{ASV}$ is a binary mapping (ACCEPT/REJECT), and $g_{ASV}$ is an extractor function to obtain the features required by the ASV. $DB$ is the database containing the voiceprints of all users. 
The system retrieves the voiceprint corresponding to the identity $U_{ID}$ from the database, extracts the features required by the ASV from both the input and the voiceprint, and outputs whether the sample $x$ belongs to $U_{ID}$.

\noindent
The CM's decision can be given as:
\begin{equation*}
    CM(x) = h_{CM}(g_{CM}(x)),
\end{equation*}

\noindent
where $h_{CM}$ is a binary mapping, and $g_{CM}$ is a feature extractor. The system extracts the relevant features from $x$ and outputs whether $x$ is spoofed or bonafide.

When an attacker attempts to impersonate a user $U_{ID}$, they must not only bypass the ASV and CM, but the content of the provided sample must match a passphrase $t$ given by the system. We assume a Speech-To-Text validation unit ($VS$) at the server that takes $t$ and the provided sample $x$ as inputs, and indicates whether this condition holds.

To capture the stealthiness to human listeners, we model these as $HJ(x,\; U_{ID},\; t)$, which takes the claimed identity, the sample provided, and the expected text as inputs. The human judge is queried when the voice authentication system's decision is challenged and serves to decide whether, indeed, the sample comes from the claimed user, is natural-sounding, and embeds the expected textual content. 

The attacker's task is to evade all these defenses. We let $\mathcal{A}$ denote the attacker's strategy used to craft the malicious samples. The inputs to $\mathcal{A}$ are $t$ and $U_{ID}$, and the output, denoted as $\mathcal{A}(t,\; U_{ID})$, is an audio sample. For the attacker's strategy to succeed, the following condition must hold:
\vspace{-1mm}
\begin{align*}
    True = & ASV(\mathcal{A}(t,\; U_{ID}),\; U_{ID}) \land CM(\mathcal{A}(t,\; U_{ID})) \\ &\land VS(\mathcal{A}(t,\; U_{ID}),\; t) \land HJ(\mathcal{A}(t,\; U_{ID}),\; U_{ID},\; t).
\end{align*}


\subsection{Attack Overview}
\label{subsec:philosophy}

As explained in \S\ref{sec:Background} and \S\ref{sec:Threat}, the only classes of attacks that may be successful against text-independent or text-prompted, combined ASV-CM systems are voice conversion (VC) and speech synthesis (SS) attacks. However, CMs may significantly degrade the potential of these attacks. The question motivating this research is whether SS and VC attacks can be optimized to defeat CM systems.

\begin{figure}[!htb]
\begin{adjustwidth}{-2cm}{}

\resizebox{0.72\textwidth}{!}{\begin{minipage}{0.72\textwidth}

    \centering
    \includegraphics[width=0.72\textwidth]{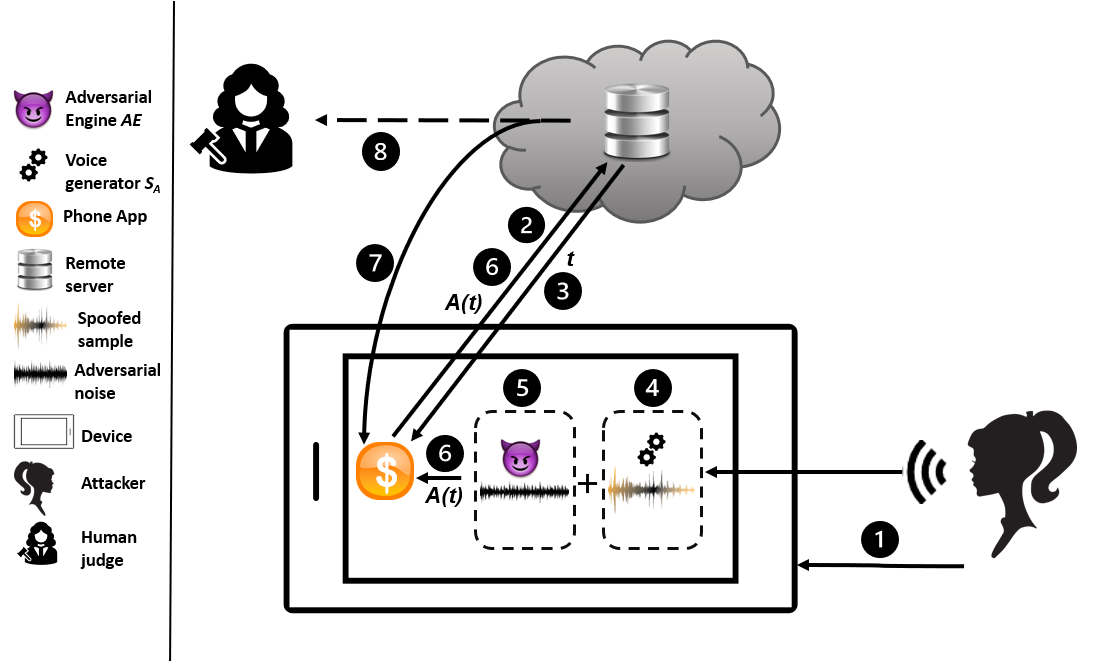}
    \caption{Attack Overview}
    \label{fig:attack}

\end{minipage}}
\end{adjustwidth}
\vspace{-1mm}
\end{figure}

State-of-the-art VC or SS algorithms~\cite{shen2018natural} can generate immaculate synthetic speech. Our hypothesis is that adversarial examples, when crafted from the outputs of these algorithms, are perfect attack candidates. 
The attacks we present target CMs via adversarial perturbations that optimize machine-generated speech bearing the victim's voiceprint. 

Given such an SS or VC algorithm denoted by $S_A$, the attack we propose is depicted in Figure~\ref{fig:attack} for the designated-app scenario:
1) The attacker accesses the app provided by the remote service on their phone and issues a command requesting a resource or triggering a transaction. The command may be a voice command, generated using $S_A$ in the victim's voice (and adversarially optimized using the same method we describe below). 2) The remote service receives the request/transaction through the app and 3) prompts the attacker with a random authentication phrase $t$. 4) The attacker generates the requested phrase in the victim's voice via $S_A$, guaranteeing acceptance by $HJ$ and $VS$. 5) The result is forwarded to the adversarial examples engine $AE$, which perturbs it slightly to generate a sample with identical content in terms of its text and the way it sounds to a human listener, while also preserving the voiceprint of the victim, but adversarially optimizing it for acceptance by the CM. 6) The adversarial example is injected into the app, which transfers it to the remote server. 7) The server verifies that the sample bears the victim's voiceprint (ASV accepts), that the textual content of the sample is $t$ ($VS$ accepts), and that the sample comes from a human (CM accepts). 
8) If the request/transaction is detected or reported as suspicious, the human judge is fooled by the seemingly genuine sample and confirms the request/transaction's authenticity. 

The attack for the over-telephony-network scenario is analogous. The only difference is that the designated service app is replaced with the phone-call app and that in step (2) the attacker has to listen to the phrase to be repeated over the phone conversation (instead of being prompted with it). The sample provided by the attacker in step (5) is forwarded to the remote server over the telephony network and needs to be encoded. The challenge the attacker faces here is the telephony network being noisy, which requires specialised techniques to robustify the attacks (see \S\ref{sec:Over-Telephony}). We assume the attacker interacts with an automated system on the other side of the phone call, rather than a human agent. Many companies deploy automated systems~\cite{zoom}---this makes our attacks easier, as VC or SS systems cannot necessarily generate high-quality speech throughout an entire conversation with a human agent.


\subsection{Fallacies of Existing Attacks}
\label{subsec:existing-attacks}

Several works have considered adversarial attacks against ASVs~\cite{kreuk2018fooling, li2020adversarial, chen2019real, jati2021adversarial, du2020sirenattack}. These attacks demonstrate how an ASV can be adversarially manipulated to accept a sample from the attacker and classify it as if it belonged to the victim. These attacks do not achieve the attacker's goal under our threat model; they are not stealthy and a human judge will be able to tell samples belonging to imposters apart. Therefore, we deem the only suitable previous attacks to be those targeting CMs. The remainder of this section scrutinizes these existing adversarial attacks against CMs~\cite{liu2019adversarial,zhang2020black}, 
and points out their limitations. Our approach, which we present afterward, introduces a paradigm shift through systematic changes to the known adversarial examples generation strategies against CMs, bypassing the limitations, and proving that adversarial attacks on CM-equipped voice authentication systems are a realistic threat. 

\subsubsection{The attacks are performed in the feature domain, not in the time domain}
\label{subsubsec:feature_space_challenge}
Assume an ML model $M:W \ \rightarrow \{0,\; 1\}$ (W is the input domain), and a sample $w \in W$, such that $M(w) = a \in \{0,\; 1\}$. An adversarial examples algorithm $\mathbb{A}$ takes $M$ and $w$, and outputs a sample $\mathbb{A}(M, w) = \hat{w}$, s.t $M(\hat{w}) = \lnot a \land \|w-\hat{w}\|\leq \epsilon$ for some value of $\epsilon$, which represents the maximal allowed perturbation (the outputted sample is indistinguishable from the original, yet it causes the model to flip its decision). Given that $M$ is a gradient-based classifier, the optimization can generally be easily solved via gradient descent/ascent techniques \cite{intriguing}. Output samples of this optimization are in the input domain of $M$. 

Our novel insight, which differentiates our work from previous attempts to attack CMs~\cite{liu2019adversarial,zhang2020black} is that this strategy fails when the target system is a CM (as we show in \S\ref{subsec:fooling_the_machine}), although it may still be optimal against a variety of other systems, including ASVs. 
Below we present our reasoning behind this previously overlooked phenomenon.


Typically, the inputs to ASVs and CMs are spectral features, which are often irreversible (e.g., MFCC). Thus, the adversarial changes are applied to spectral representations of audio signals, outputting adversarial spectrograms. To transmit a signal to the remote ASV or CM, the time-domain signal has to be restored from the modified spectral representations, which poses two major challenges:


First, note that the process of synthesizing waveforms from spectrograms is an integral part of numerous VC and SS algorithms (WaveNet-based approaches~\cite{oord2016wavenet}). Hence, by nature, this process introduces fingerprints that are detectable by CMs  trained to spot these algorithms. In other words, the restoration procedure that occurs after the optimization reintroduces machine artefacts that the adversarial optimization tried to eliminate, nullifying the effects of the attack.

Second, a spectral representation consists of two components: the phase and magnitude. ASVs and CMs discard the phase when processing the signal. However, the phase is crucial for recovering an audio waveform from a spectrogram. This is the case when the waveform is restored via the Griffin-Lim~\cite{gl} algorithm, as done in the previous works~\cite{li2020adversarial, liu2019adversarial, zhang2020black}. Other modern approaches (to restore waveforms), such as WaveNet vocoders~\cite{oord2016wavenet}, still suffer from the above first limitation. Furthermore, these may not always be applicable due to the discrepancies in the features and their hyper-parameters between those required for vocoders and what is used by ASVs and CMs. The above works resort to the following technique: 
the spectral representation is first extracted from the original audio, from which the attacker obtains the magnitude and the phase. Afterward, the magnitude is adversarially perturbed. 
The restoration process takes place using the modified magnitude and the original phase. Importantly, the phase used for restoring the signal does not match the modified sample as it belongs to the original recording. This approach is unrealistically optimistic since CMs are trained to detect machine-generated speech and to spot machine artifacts. Signals generated using mismatching magnitudes and phases cannot come from humans. This hurdle is critical when attacking a CM, while ASVs may not be as robust. Attacks against ASVs remain highly successful when mounted in the feature space even when the waveform is reconstructed as described above~\cite{li2020adversarial, kreuk2018fooling}. Our insights lead us to abandon the traditional method adopted by all the existing attacks against CMs~\cite{liu2019adversarial, zhang2020black} 
and serve as the basis for the strategies we present later. 

Despite the above, existing attacks against CMs~\cite{liu2019adversarial,zhang2020black} 
report high success rates. We argue that this is because they do not reconstruct audio from the perturbed spectrograms and assume that these can be fed directly into the attacked server. However, in realistic settings, the attacker may only be able to forward a waveform to the server, which will extract the spectrogram independently. As we show in \S\ref{subsec:fooling_the_machine}, synthesizing speech from the adversarial spectrograms and  forwarding it to the server severely diminishes these attacks' success rates.

\subsubsection{The attacks do not consider a combined ASV-CM deployment}
Existing attacks on CMs~\cite{liu2019adversarial, zhang2020black} 
do not consider an ASV in their evaluation. The presented attacks perturb spoofed samples that are initially rejected by a CM and re-evaluate the CM's performance on the perturbed samples. The reported success rates only demonstrate the number of samples successfully optimized to circumvent the CM. Yet, during this optimization, certain artefacts are introduced into the samples that can cause them to be rejected by an ASV.

\subsubsection{The attacks are performed in a relaxed black-box setting} While existing attacks on CMs~\cite{liu2019adversarial, zhang2020black} 
consider a somewhat black-box model by varying the architecture of the target CMs, crafting examples against shadow models, and exploiting the transferability property to mount successful attacks, the dataset on which all the models are trained is identical. In such cases, it is only natural that a larger portion of the adversarial examples would transfer. In realistic scenarios, the attacker typically does not have access to the same dataset the target system was trained on. When evaluating the attacks under this restriction, we expect a significant degradation in the transferability rates.

\subsection{Our Attack} 
\label{subsec:our_attack}
\subsubsection{Overview}

We present the details of our attack and explain the mathematical formulation of our adversarial examples generation process to craft high-quality spoofed speech.

Our attack consists of two components: a spoofed speech generator $S_A$ and an adversarial examples engine $AE$ that optimizes a spoofed speech sample to make it bypass the CM. 
We redefine the attacker's algorithm $\mathcal{A}$ introduced earlier as follows: $\mathcal{A} = AE \circ S_A$. We assume $S_A$ is a state-of-the-art SS or VC algorithm, and shift our focus to the optimizer component $AE$, which is where our technical contribution lies.

$AE$ is defined as an algorithm $AE: X \rightarrow X$, which given an input voice recording $x$, generates a recording $x_{adv}$, such that $x_{adv}$ is the solution to the following optimization problem:

\vspace{-3mm}

\begin{equation*}
\begin{split}
     &argmin_{x^\prime}\; \ell(x^\prime) \\
    &s.t\; \| x - x^\prime \|_\infty \leq \epsilon.
\end{split}
\end{equation*}

$\ell$ is a loss function to be discussed later, constructed by the attacker based on the outputs of a locally-trained shadow model to which they have white-box access. The algorithm strives to optimize a sample $x$ by outputting a sample $x_{adv}$, which 1) minimizes the loss function of the shadow model, which corresponds to the sample's ability to bypass a CM, and 2) ensures the norm of the difference between the original and adversarial example is below some small threshold $\epsilon$. 

\subsubsection{Loss Function}
\label{sec:our_attack:joint_loss_function}

We next introduce our loss function $\ell$. Recall that our attacker wishes to perturb the input spoofed voice sample to bypass a CM while preserving the victim's voiceprint in the sample, the textual content, and the resemblance to the victim's voice to human listeners. The requirements to preserve the resemblance to the victim's voice (as perceived by humans and machines---recall that ASVs are vulnerable to spoofing attacks) and the textual content are met by choosing a small enough $\epsilon$. Hence, the loss function is only concerned with bypassing the CM. 

The process of constructing this loss function is guided by our insights from \S\ref{subsec:existing-attacks}, leading to two requirements an $AE$ must satisfy to craft successful adversarial examples:

\begin{enumerate} [i.]
    \item The perturbations must withstand the effects of lossy (inverse) spectral transformations.
    \item The victim’s voiceprint must be preserved.
\end{enumerate}

Below, we address the two requirements separately and present loss function candidates.

\begin{enumerate} [i.] 
    \item \textbf{Adversarial Manipulation of Time-Domain Signals}
    \end{enumerate}
\label{subsubsec:raw_manip}
To circumvent the hurdles arising when introducing adversarial perturbations to spectrograms and reconstructing signals from them, we propose generating the adversarials in the input (i.e., time) domain, directly perturbing the input waveforms. 
Suppose a loss function $\ell_F: F \rightarrow \mathbb{R}$, where $F$ is the feature space. Since ASVs and CMs operate on spectral representations, such loss functions are the most common. When generating our adversarials, we wish to optimize the loss w.r.t the raw waveform itself. Given the loss function $\ell_F$ defined over the set of features extracted from an input, and the feature extractor $g: X \rightarrow F$, we define the loss $\ell$ as $\ell = \ell_F \circ g$. The assumption is that there is a differentiable $\ell_F$ with a gradient-supported implementation. On the other hand, 
feature extractors for features such as MFCC, LPMS, or CQCC are typically implemented in libraries that do not support gradient back-propagation (e.g., librosa), since such functionality is not needed when learning the weights of a model whose inputs are the trainable parameters and not the signals. 
Since the typical procedure to perform adversarial optimization involves multiple steps of gradient descent (ascent) w.r.t the input, we need a method to propagate the gradients of a feature extractor back to the raw waveform stage. 

To solve this, we re-implement the feature extractors with gradient support in Python's Tensorflow. 
Our feature extractors can therefore receive and operate on Tensors constructed from our input vectors. These functions are capable of calculating the gradients of the features w.r.t the time-domain input signal. 

Finally, using the chain rule, we are now able to propagate the gradients back to the raw signal layer, since:
\vspace{-1mm}
\begin{equation*}
    \ell = \ell_F \circ g \implies \ell(x) = \ell(g(x)) \implies \frac{\partial \ell}{\partial x} = \frac{\partial \ell_F}{\partial g} \cdot \frac{\partial g}{\partial x}.
    \label{eq:chain}
\end{equation*}
\vspace{-2.5mm}

This approach eliminates all the problems raised in \S\ref{subsubsec:feature_space_challenge} as the adversarial perturbations are applied directly to the input, preventing the loss in the inverse transformation stage.
\\
\vspace{-4mm}

\begin{algorithm}[t]
\SetAlgoLined
\SetKwInOut{KwIn}{Input}
\SetKwInOut{KwOut}{Output}
\SetKwFunction{Initialize}{Initialize}
 \KwIn{A shadow CM model $CM_S$, the gradient-supporting feature extractor for $CM_S$, $g^S_{CM}$, and a spoofed time-domain voice sample $x$}

\KwOut{Optimized adversarial spoofed sample $x^\prime$}

\Initialize: $\delta \sim U(-\epsilon, \epsilon)$, $x^\prime = x + \delta$

 \For{$t = 0,\; 1,...$}{
     $feats = g^S_{CM} (x^\prime)$ \\
     $Loss = CM_S (feats)$ \\
     $Grads = \nabla_{x^\prime} (\nabla_{feats} (Loss) * feats)$ \\
     $x^\prime \longleftarrow clip_{\epsilon(x)}\{x^\prime - \alpha * sign(Grads)\}$\\
 }
  $return\; x^\prime$
 \caption{ADVCM}
 \label{alg:ADVCM}
\end{algorithm}

\textbf{Proposed loss 1.} \textit{ADVCM} (Adversarial CM): Assuming the attacker can build a shadow CM, denoted as $CM_S$, 
the straightforward approach is to use this model's loss function. By minimizing this loss, we guarantee that our shadow CM is more likely to accept this sample (we assume a lower loss corresponds to a bonafide sample). The transferability of adversarial examples~\cite{wu2020towards} will be responsible for this sample bypassing the target CM as well. This method already addresses the challenges in \textbf{all} prior works on adversarial examples against CMs, since our method of directly manipulating time-domain signals eliminates reconstruction errors. This approach is presented in Algorithm~\ref{alg:ADVCM}. 

The algorithm initializes a random noise $\delta$ as the adversarial perturbation, where we sample $\delta$ uniformly from $[-\epsilon,\; \epsilon]$. We solve the optimization problem by performing gradient descent to minimize the loss function. To avoid needing to reconstruct the signal from the spectral representation, we operate on $x$ as a raw signal. In each iteration, we take the current sample $x^\prime$ and extract the features used by our shadow CM via our gradient-supported feature extractor:

\vspace{-3mm}
\begin{equation*}
    feats = g^S_{CM} (x^\prime).
\end{equation*}

The features are used to calculate the CM's loss as $Loss = CM_S (feats)$. The chain rule is used to calculate the gradients:
\vspace{-1.5mm}
\begin{equation*}
    Grads = \nabla_{x^\prime} (\nabla_{feats} (Loss) * feats).
\end{equation*}

\noindent
The gradients enable us to perform a gradient descent step
\vspace{-1.5mm}
\begin{equation*}
x^\prime \longleftarrow clip_{\epsilon(x)}\{x^\prime - \alpha * sign(Grads)\}
\end{equation*}
 to converge to an optimal solution. Here, $\alpha$ is a learning rate we adjust empirically, and $\epsilon(x)$ represents the environment $(x-\epsilon, x + \epsilon)$, so that at every step the sample will stay in the $L_\infty$ environment of $x$ to meet the minimal distortion requirement. The above procedure returns an adversarial time-domain sample $x_{adv}$.
 \\
 
 \label{sec:ADVSR}
\vspace{-2mm} 
\begin{enumerate} [ii.] 
    \item \textbf{Preserving The Victim's Voiceprint}
    \end{enumerate}
 While ADVCM may successfully defeat any CM model in the white-box setting, this is not the case for combined ASV-CM deployments. 
An optimization that ignores the ASV may fail to craft adversarials that successfully bypass both systems. Generating adversarial examples entails applying perturbations to spoofed samples. Regardless of whether these samples are initially capable of passing an ASV, the added noise may ``remove'' the speaker's voiceprint. We refer to this as the CM overfitting phenomenon---the perturbations applied may indeed be useful for bypassing spoofing detectors by eliminating machine artefacts. However, the shadow CMs used to apply these changes are not trained to recognize or retain the specific user's voiceprint. Therefore, during optimization these user specific telltales may be tainted since the CM is mainly concerned with eliminating machine signatures, which leads to the samples being rejected by the target ASV. 

Given these observations, an ideal adversarial examples generation strategy would entirely mitigate the CM overfitting effect to ensure robustness. Therefore, the reliance on a shadow CM is sub-optimal, and we present a superior alternative next.

\begin{algorithm}[t]
\SetAlgoLined
\SetKwInOut{KwIn}{Input}
\SetKwInOut{KwOut}{Output}
\SetKwFunction{Initialize}{Initialize}
 \KwIn{A shadow ASV model $ASV_S$, the gradient-supporting feature extractor for $ASV_S$, $g^S_{ASV}$, a bonafide sample $y$ belonging to some user $U$, and a spoofed time-domain voice sample $x$ generated for the same user $U$}

\KwOut{Optimized adversarial spoofed sample $x^\prime$}

\Initialize: $\delta \sim U(-\epsilon, \epsilon)$, $x^\prime = x + \delta$

 \For{$t = 0,\; 1,...$}{
     $feats_x = g^S_{ASV} (x^\prime)$ 
     \\
     $feats_y = g^S_{ASV} (y)$ \\
     $Loss = ASV_S (feats_x,\; feats_y)$ \\
     $Grads = \nabla_{x^\prime} (\nabla_{feats_x} (Loss) * feats_x)$ \\
     $x^\prime \longleftarrow clip_{x,\; \epsilon}\{x^\prime - \alpha * sign(Grads)\}$\\
 }
  $return\; x^\prime$
 \caption{ADVSR}
 \label{alg:ADVSR}

\end{algorithm}

\textbf{Proposed loss 2.} \textit{ADVSR} (Adversarial speaker regularization): We propose employing a novel loss function that when minimized leads indirectly to spoofed examples of higher quality that can bypass CMs without ``removing'' the victim's voiceprint. Our loss of choice is that of a shadow ASV. Below we explain why such a function is capable of producing universal adversarials that defeat both CMs and ASVs. Consistent with the ASVs we experiment with, we assume the victim's voiceprint is a bonafide recording in their voice. 
Given that the attacker has access to a shadow ASV, denoted as $ASV_S$, whose loss is $\ell_{ASV_S}$, we argue that if the attacker obtains a {\it single} bonafide reference recording of the victim (which can be different from the victim's voiceprint stored on the server), denoted as $y$, they can optimize a spoofed sample $x$ via the loss function:
\begin{equation*}
     \ell_{ASV_S}(x, y).
\end{equation*}

Interestingly this will lead to a more natural sample capable of bypassing CMs, while retaining the voiceprint. Our hypothesis is as follows:  
The spoofed sample is generated by an SS or VC algorithm that attempts to mimic the victim's voiceprint. Since these algorithms have pitfalls, they produce artefacts; the same artefacts that help CMs spot spoofed samples. 
These nuances might manifest themselves through machine-like effects that humans generally cannot produce, or in the form of absent human characteristics computers fail to capture. In both cases, the differences in our samples testify to the involvement of a machine in the creation of the spoofed sample, which will therefore exhibit differences from \textbf{any bonafide sample}. 

By optimizing the spoofed sample based on the ASV's loss, the above differences are eliminated. The idea is to ``engrave" the user's voiceprint into the spoofed sample and further optimize it by including the ASV in the process. The rationale we follow is that the ASV's loss indicates that the speakers in the two samples $x$ and $y$ are different, and we attribute that to one of the samples being ``spoken" by the machine. When the ASV eliminates these differences, the machine characteristics are removed as a byproduct. The sample $x$ is, hence, less machine-like and is capable of bypassing the CM, without using a shadow CM for the optimization. Another advantage of our novel strategy is that the ASV retains the victim's voiceprint. 
Finally, by integrating the time-domain adversarial perturbation approach introduced earlier, we avoid destructive reconstruction noises, which is key to making these attacks succeed. Algorithm~\ref{alg:ADVSR} describes this approach. Overall, it is identical to Algorithm~\ref{alg:ADVCM}, except that the loss function is replaced by a shadow ASV's loss.

\textbf{Proposed loss 3.} \textit{ADVJOINT}: Finally, we propose a joint loss function that combines the ASV loss and the CM loss. The ASV loss serves as a regularizer that prevents the adversarial examples from overfitting to CMs. The CM loss strives to provide more robust task-specific perturbations that ADVSR may miss. 
The capability of the ASV loss to properly induce the desired perturbations sufficient to fool the CM may still be limited due to the task differences and is also influenced by the extent to which the ASV and the CM are affected by similar low-level characteristics.
Jointly, we expect the CM loss to produce the desired perturbations while the ASV prevents the above overfitting and preserves the victim's voiceprint. It is worth noting that defining and using a joint loss function is only made possible thanks to the ability of our approach to propagate the error of CMs to the input in the time domain, since ASVs and CMs rely on different feature sets. Algorithm~\ref{alg:ADVJOINT} describes this approach.  The joint loss function the optimization stage aims to minimize is
\begin{equation*}
    \lambda_1 \ell_{ASV_S}(x,\; y) + \lambda_2 \ell_{CM_S}(x),
\end{equation*}

\noindent
where $y$ is a bonafide sample, and $\lambda_1$ and $\lambda_2$ are two hyperparameters to be tuned.

\begin{algorithm}[]
\SetAlgoLined
\SetKwInOut{KwIn}{Input}
\SetKwInOut{KwOut}{Output}
\SetKwFunction{Initialize}{Initialize}
 \KwIn{A shadow ASV model $ASV_S$, The gradient-supporting feature extractor for $ASV_S$, $g^S_{ASV}$, A shadow CM model $CM_S$, The gradient-supporting feature extractor for $CM_S$, $g^S_{CM}$, a bonafide sample $y$ belonging to some user $U$, and a spoofed time-domain voice sample $x$ generated for the same user $U$}

\KwOut{Optimized adversarial spoofed sample $x^\prime$}

\Initialize: $\delta \sim U(-\epsilon, \epsilon)$, $x^\prime = x + \delta$

 \For{$t = 0,\; 1,...$}{
     $feats_x = g^S_{ASV} (x^\prime)$ \\
     $feats_y = g^S_{ASV} (y)$ \\
     $Loss_{ASV} = ASV_S (feats_x,\; feats_y)$ \\
     $Grads_{ASV} = \nabla_{x^\prime} (\nabla_{feats_x} (Loss) * feats_x)$ \\
     $feats = g^S_{CM} (x^\prime)$ \\
     $Loss_{CM} = CM_S (feats)$ \\
     $Grads_{CM} = \nabla_{x^\prime} (\nabla_{feats} (Loss) * feats)$ \\
     $Grads = \lambda_1 Grads_{CM} + \lambda_2 Grads_{ASV}$\\
     $x^\prime \longleftarrow clip_{x,\; \epsilon}\{x^\prime - \alpha * sign(Grads)\}$\\
 }
  $return\; x^\prime$
 \caption{ADVJOINT}
 \label{alg:ADVJOINT}
\end{algorithm}

\section{Experiments}
Our experiments take place in the black-box setting. We assume the attacker has no access to the datasets used to train the target models, nor to the architectures of these systems. For reference, and to demonstrate the effects of the transferability property on the success rate of the attacks, we provide results for the white-box setting as well. Following our threat model, we evaluate our attacks w.r.t three factors: 1) Ability to fool a machine (i.e., ASVs and CMs), 2) Content preservation, and 3) Indistinguishability from genuine samples to a human judge. 

\subsection{Fooling The Machine}
\label{subsec:fooling_the_machine}
We first assess our attacks' ability to fool joint ASV-CM systems. 
We assume the attacker has a state-of-the-art VC or SS algorithm (see \S\ref{subsec:our_attack}) and deem evaluating the quality of the generated spoofed speech out of scope. Thus, our starting point is a spoofed sample, to which we add adversarial noise.

\textbf{Dataset.} We mainly use the Logical Access (see \S\ref{sec:RelatedWork}) portion of the ASVspoof2019 dataset~\cite{wang2020asvspoof}, which includes samples from 107 speakers (46 males and 61 females). For each speaker, the dataset contains bonafide samples and samples spoofed in the speaker's voice generated using VC or SS algorithms. The dataset is partitioned into three sections: training, development, and evaluation. Each of the training and development datasets contain samples from 20 speakers, while the evaluation dataset's samples come from 67 speakers. The datasets are disjoint in terms of the included speakers, which is consistent with our threat model where we assume the attacker may not have access to the target models' internals. 

\textbf{Settings.} We first present experiments for the designated-app scenario. Experiments for the over-telephony-network scenario are in \S\ref{sec:Over-Telephony}. 
For CMs, we experiment with three different architectures: $lcnnHalf$, $lcnnFull$, and $SENet$~\cite{liu2019adversarial}. These, or highly similar architectures, were submitted to the ASVspoof2019 challenge and scored among the best 15 competing systems. For the ASVs, we adopt the GMM i-vector architecture by Li et al.~\cite{li2020adversarial}. We use different shadow/target architectures by varying the hyperparameters that determine the dimensionality of the i-vector (shadow: 400, target: 300), and the number of the Gaussians in the GMM model (shadow: 2048, target: 1024). We argue that these variations sufficiently capture real-world attackers' knowledge even in the black-box scenario, since GMM i-vector models are widely used and so it is likely that it would be the choice of both the attacker and the target ASV. Due to the large-scale nature of our study and the extensive computing resources required, we choose not to explore x-vector architectures here.  However, the experiments in~\S\ref{sec:Over-Telephony} assume a more restrictive scenario (over-telephony-network) and the high success rates presented there against target systems with x-vector ASVs (using i-vector shadows) guarantee success under the scenario we consider here as well.

The input features to our CMs are LPMSs, while to the ASVs the inputs are MFCCs. For the target system, we deploy an ASV and a CM model that jointly judge a given sample. On the attacker's side, we install shadow models based on the method used for the attack (only a shadow CM for ADVCM and CMSPEC, only a shadow ASV for ADVSR, and both a shadow CM and a shadow ASV for ADVJOINT). CMSPEC is the method of generating adversarial examples in the feature domain, as done by existing attacks~\cite{liu2019adversarial,zhang2020black}.

We train the shadow CMs ($lcnnHalf_s$, $lcnnFull_s$, and $SENet_s$) on the ASVspoof2019 training section, while the target CMs ($lcnnHalf_t$, $lcnnFull_t$, and $SENet_t$) are trained on the development section. This way, our black-box assumptions are preserved. The VC and SS algorithms used to generate the spoofing examples included in the training and development sections were the same, which as stated in~\S\ref{subsec:attacker_knowledge} still results in an attacker only as powerful as a real-world adversary. To train the ASVs, we use only bonafide examples. The shadow ASV $asv_s$ is trained using the ASVspoof2019 training section, while the target ASV $asv_t$ uses the development section. The baseline performance of all models is in Table~\ref{table:eer} in Appendix~\ref{app:models}. All experiments run on a server with a Gentoo 2.6 OS on an Intel Xeon Gold 6238 Cascade Lake 2.1 GHz CPU, with 64G RAM and two Turing T4 GPUs with 16G of memory each.

\textbf{Experiment Design.} 
Since we have three different CM architectures and a single ASV, we have three options for the combined target system. Similarly, we end up with three possible combined shadow systems. Since the target architecture is not known to the attacker, we do not include the same architecture for both the shadow and target systems in the same experiment. Hence, we in total have six attack configurations. In all these configurations, the shadow and target systems have different architectures and different training sets. 

\textbf{Evaluation Metrics.} We are concerned with instances where an attacker attempts to fool a voice authentication system with a fake sample. Hence, we focus on spoofed speech recordings only and do not include any bonafide samples in our evaluation. Specifically, we randomly select 20K samples from the $\sim$ 63K spoofed speech instances in the evaluation section of ASVspoof2019, and define our attack's success rate to be the False Acceptance Rate (FAR) on the adversarials generated from these samples using one of our algorithms. For an adversarial sample $A$ to be accepted, it needs to bypass both an ASV and a CM. For each $A$, the target system receives the speaker $ID$ to whom $A$ supposedly belongs and retrieves the enrollment utterance of the same speaker stored in the system. This enrollment phrase is one of the bonafide recordings by that speaker from the ASVspoof2019 evaluation section. By doing so, the target system can invoke the ASV model on $A$ to verify the claimed identity based on the retrieved voiceprint.

We use the following terminology: $X/Y$ represents an ASV-CM deployment (either both are shadow models or both are target systems), with $X$ being the ASV and $Y$ being the CM. $S-T$ represents a shadow-target configuration. That is, we craft adversarial examples using shadow system $S$ to attack  target system $T$. Naturally, when reporting results in the white-box setting, the success rate is only w.r.t the shadow system present; for CMSPEC and ADVCM, the white-box results represent the samples' ability to bypass the shadow CM only (since there is no shadow ASV). For ADVSR, the white-box results similarly represent the shadow ASV's confidence in the sample matching the victim user's voiceprint. For ADVJOINT, the results take into account a combined shadow ASV and CM. Trivially, all black-box systems---with the exception of CMSPEC (see below)---consist of both an ASV and a CM.

\begin{table}[]

\resizebox{0.5\textwidth}{!}{\begin{tabular}{l|l|l|l|l|l|l}
\hline
\textbf{{\backslashbox{\textit{Shadow}}{\textit{epsilon}}}} & \textit{\textbf{$\epsilon$ = 0}} & \multicolumn{1}{c|}{\textit{\textbf{\begin{tabular}[c]{@{}c@{}}$\epsilon$\\  =0.1\end{tabular}}}} & \multicolumn{1}{c|}{\textit{\textbf{\begin{tabular}[c]{@{}c@{}}$\epsilon$\\  =1\end{tabular}}}} & \multicolumn{1}{c|}{\textit{\textbf{\begin{tabular}[c]{@{}c@{}}$\epsilon$\\  =5\end{tabular}}}} & \multicolumn{1}{c|}{\textit{\textbf{\begin{tabular}[c]{@{}c@{}}$\epsilon$\\  =10\end{tabular}}}} & \multicolumn{1}{c}{\textbf{\begin{tabular}[c]{@{}c@{}}$\epsilon$\\  =20\end{tabular}}} \\ \hline
\textbf{$SENet_s$ (direct)}                     & 8.58\%                           & 20.99\%                                                                                           & 74.34\%                                                                                         & 33.64\%                                                                                         & \textbf{100.00\%}                                                                                & 99.02\%                                                                                \\
\textbf{$SENet_s$ (reconstructed)}              & N/A                              & 10.88\%                                                                                           & 26.25\%                                                                                         & 3.06\%                                                                                          & \textbf{59.30\%}                                                                                 & 27.70\%                                                                                \\
\textbf{$lcnnFull_s$ (direct)}                  & 3.48\%                           & 6.71\%                                                                                            & 51.83\%                                                                                         & \textbf{100.00\%}                                                                               & \textbf{100.00\%}                                                                                & \textbf{100.00\%}                                                                      \\
\textbf{$lcnnFull_s$ (reconstructed)}           & N/A                              & 4.89\%                                                                                            & 22.56\%                                                                                         & 86.24\%                                                                                         & 99.83\%                                                                                          & \textbf{100.00\%}                                                                      \\
\textbf{$lcnnHalf_s$ (direct)}                  & 4.76\%                           & 24.81\%                                                                                           & 99.58\%                                                                                         & \textbf{100.00\%}                                                                               & \textbf{100.00\%}                                                                                & \textbf{100.00\%}                                                                      \\
\textbf{$lcnnHalf_s$ (reconstructed)}           & N/A                              & 8.45\%                                                                                            & 49.02\%                                                                                         & \textbf{72.18\%}                                                                                & 50.00\%                                                                                          & 1.91\%                                                                                 \\ \hline
\end{tabular}}
\caption{Success rate of white-box attacks in the feature domain (CMSPEC)} 
\label{tab:main_cmspec_wb}

\resizebox{0.5\textwidth}{!}{\begin{tabular}{l|l|l|l|l|l|l}
\hline

\textbf{{\backslashbox{\textit{Shdw - Tgt}}{\textit{epsilon}}}}             & \textit{\textbf{$\epsilon$ = 0}} & \multicolumn{1}{c|}{\textit{\textbf{\begin{tabular}[c]{@{}c@{}}$\epsilon$\\  =0.1\end{tabular}}}} & \multicolumn{1}{c|}{\textit{\textbf{\begin{tabular}[c]{@{}c@{}}$\epsilon$\\  =1\end{tabular}}}} & \multicolumn{1}{c|}{\textit{\textbf{\begin{tabular}[c]{@{}c@{}}$\epsilon$\\  =5\end{tabular}}}} & \multicolumn{1}{c|}{\textit{\textbf{\begin{tabular}[c]{@{}c@{}}$\epsilon$\\  =10\end{tabular}}}} & \multicolumn{1}{c}{\textbf{\begin{tabular}[c]{@{}c@{}}$\epsilon$\\  =20\end{tabular}}} \\ \hline
\textbf{$SENet_s$ - $lcnnHalf_t$}    & 8.06\% & \textbf{10.493\%}                  & 6.415\%                                                                                            & 1.656\%                                                                                          & 0.81\%                                                                                          & 1.23\%                                                                                                                                 \\
\textbf{$SENet_s$ - $lcnnFull_t$}    & 4.36\%                  & 5.1\%                                                                                            & 4.93\%                                                                                          & \textbf{6.46\%}                                                                                          & 2.97\%                                                                                           & 0.55\%                                                                                 \\
\textbf{$lcnnFull_s$ - $SENet_t$}    & 7.73\%                           & 8.49\%                                                                                            & 8.88\%                                                                                          & \textbf{9.09\%}                                                                                 & 5.69\%                                                                                           & 0.00\%                                                                                 \\
\textbf{$lcnnFull_s$ - $lcnnHalf_t$} & 8.06 & \textbf{10.535\%}                           & 9.81\%                                                                                   & 5.4\%                                                                                          & 4.04\%                                                                                          & 2.12\%                                                                                                                                        \\
\textbf{$lcnnHalf_s$ - $SENet_t$}    & 7.73\%                           & 8.453\%                                                                                   & 8.5\%                                                                                          & \textbf{9.05\%}                                                                                          & 7.69\%                                                                                           & 0.3\%                                                                                 \\
\textbf{$lcnnHalf_s$ - $lcnnFull_t$} & 4.36\%                  & 5.14\%                                                                                            & \textbf{5.31\%}                                                                                          & 4.63\%                                                                                         & 3.48\%                                                                                           & 0.47\%                                                                                 \\ \hline
\end{tabular}}
\caption{Success rate of black-box attacks in the feature domain (CMSPEC)}
\label{tab:main_cmspec}
\vspace{-5mm}
\end{table}

\begin{table*}
\centering
\resizebox{\textwidth}{!}{\begin{tabular}{@{}clllllllllllll@{}}
\toprule
                                                            & \textbf{$\epsilon$ = 0} &
                                                   \multicolumn{3}{c}{\textbf{$\epsilon$ = 0.001}} &   
                                                   \multicolumn{3}{c}{\textbf{$\epsilon$ = 0.003}} &   \multicolumn{3}{c}{\textbf{$\epsilon$ = 0.005}} &
                                                   \multicolumn{3}{c}{\textbf{$\epsilon$ = 0.007}}   \\ \midrule
{\backslashbox{\textit{Shadow - Target}}{\textit{Attack Type}}}                                       &                                  & \textbf{ADVCM}        & \multicolumn{1}{c}{\textbf{ADVSR}} & \multicolumn{1}{c}{\textbf{ADVJOINT}} & \textbf{ADVCM}        & \multicolumn{1}{c}{\textbf{ADVSR}} & \multicolumn{1}{c}{\textbf{ADVJOINT}}    & \textbf{ADVCM}        & \multicolumn{1}{c}{\textbf{ADVSR}} & \multicolumn{1}{c}{\textbf{ADVJOINT}} & \textbf{ADVCM}            & \multicolumn{1}{c}{\textbf{ADVSR}}     & \textbf{ADVJOINT}      \\ \midrule
\multicolumn{1}{c|}{\textbf{$asv_s/lcnnHalf_s$ - $asv_t/lcnnFull_t$}} & \multicolumn{1}{l|}{1.45\%}      & \multicolumn{1}{l}{6.9\%} & 11.59\%                    & \multicolumn{1}{l|}{\textbf{12.22\%}}       & \multicolumn{1}{l}{3.33\%} & 6.97\%                                & \multicolumn{1}{l|}{7.37\%}          & \multicolumn{1}{l}{2.16\%} & 3.42\%                             & \multicolumn{1}{l|}{4.96\%}       & \multicolumn{1}{l}{1.26\%}     & 1.05\%                                  & 3.06\%             \\
\multicolumn{1}{c|}{\textbf{$asv_s/lcnnHalf_s$ – $asv_t/SENet_t$}}    & \multicolumn{1}{l|}{3.64\%}      & \multicolumn{1}{l}{42.5\%} & 30.51\%                             & \multicolumn{1}{l|}{51.32\%}       & \multicolumn{1}{l}{66.58\%} & 86.39\%                             & \multicolumn{1}{l|}{84.67\%}          & \multicolumn{1}{l}{64.64\%} & 93.43\%                             & \multicolumn{1}{l|}{86.54\%}       & \multicolumn{1}{l}{61.55\%}     & \textbf{91.37\%}                        & 86.77\%             \\
\multicolumn{1}{c|}{\textbf{$asv_s/lcnnFull_s$ – $asv_t/lcnnHalf_t$}} & \multicolumn{1}{l|}{3.9\%}      & \multicolumn{1}{l}{24.4\%} & 23.06\%                             & \multicolumn{1}{l|}{\textbf{37.29\%}}       & \multicolumn{1}{l}{7.12\%} & 29.3\%                             & \multicolumn{1}{l|}{14.4\%}          & \multicolumn{1}{l}{5.01\%} & 36.1\%                    & \multicolumn{1}{l|}{8.47\%}       & \multicolumn{1}{l}{5.15\%}     & 28.12\%                                 & 7.63\%             \\
\multicolumn{1}{c|}{\textbf{$asv_s/lcnnFull_s$ – $asv_t/SENet_t$}}    & \multicolumn{1}{l|}{3.64\%}      & \multicolumn{1}{l}{26.37\%} & 30.51\%                             & \multicolumn{1}{l|}{41.1\%}       & \multicolumn{1}{l}{39.99\%} & 86.39\%                             & \multicolumn{1}{l|}{67.8\%}          & \multicolumn{1}{l}{56.89\%} & 93.43\%                             & \multicolumn{1}{l|}{85.86\%}       & \multicolumn{1}{l}{61.8\%}     & \textbf{93.57\%}                        & 91.49\%             \\
\multicolumn{1}{c|}{\textbf{$asv_s/SENet_s$ – $asv_t/lcnnHalf_t$}}    & \multicolumn{1}{l|}{3.9\%}      & \multicolumn{1}{l}{35.33\%} & 23.06\%                             & \multicolumn{1}{l|}{46.3\%}       & \multicolumn{1}{l}{50.43\%} & 29.29\%                             & \multicolumn{1}{l|}{64.81\%}          & \multicolumn{1}{l}{52.24\%} & 36.1\%                             & \multicolumn{1}{l|}{72.37\%}        & \multicolumn{1}{l}{50.92\%}     & 28.12\%                                 & \textbf{73.5\%}    \\
\multicolumn{1}{c|}{\textbf{$asv_s/SENet_s$ – $asv_t/lcnnFull_t$}}    & \multicolumn{1}{l|}{1.45\%}      & \multicolumn{1}{l}{22.73\%} & 11.59\%                             & \multicolumn{1}{l|}{\textbf{34.09\%}}       & \multicolumn{1}{l}{15.26\%} & 6.97\%                                & \multicolumn{1}{l|}{26.77\%} & \multicolumn{1}{l}{5.24\%} & 3.42\%                             & \multicolumn{1}{l|}{11.91\%}       & \multicolumn{1}{l}{1.97\%}     & 1.05\%                                  & 4.6\%             \\ \bottomrule
\end{tabular}}
\caption{Success rate of our attacks for different values of $\epsilon$ in the black-box setting}
\label{tab:main_exp}
\vspace{-4mm}
\end{table*}

\begin{table}[t]
\centering
\resizebox{0.5\textwidth}{!}{\begin{tabular}{llllll}
\hline
\multicolumn{1}{l|}{\textbf{}}              & \textbf{$\epsilon$ = 0} & \multicolumn{1}{c}{\textbf{\begin{tabular}[c]{@{}c@{}}$\epsilon$\\  =0.001\end{tabular}}} & \multicolumn{1}{c}{\textbf{\begin{tabular}[c]{@{}c@{}}$\epsilon$\\  =0.003\end{tabular}}} & \multicolumn{1}{c}{\textbf{\begin{tabular}[c]{@{}c@{}}$\epsilon$\\  =0.005\end{tabular}}} & \multicolumn{1}{c}{\textbf{\begin{tabular}[c]{@{}c@{}}$\epsilon$\\  =0.007\end{tabular}}} \\ \hline
\multicolumn{6}{c}{\textbf{Whitebox-ADVSR}}                                                                                                                                                                                                                                                                                                                                                                                                                                                       \\ \hline
\multicolumn{1}{l|}{\textbf{$asv_s$}}          & 69.82\%                          & 100.00\%                                                                                           & 100.00\%                                                                                          & 100.00\%                                                                                            & 100.00\%                                                                                           \\ \hline
\multicolumn{6}{c}{\textbf{Whitebox-ADVCM}}                                                                                                                                                                                                                                                                                                                                                                                                                                                           \\ \hline
\multicolumn{1}{l|}{\textbf{$SENet_s$}}         & 10.85\%                          & 99.99\%                                                                                            & 100.00\%                                                                                          & 100.00\%                                                                                            & 100.00\%                                                                                           \\
\multicolumn{1}{l|}{\textbf{$lcnnFull_s$}}      & 5.19\%                           & 99.99\%                                                                                            & 100.00\%                                                                                          & 100.00\%                                                                                            & 100.00\%                                                                                           \\
\multicolumn{1}{l|}{\textbf{$lcnnHalf_s$}}      & 6.22\%                           & 99.99\%                                                                                            & 100.00\%                                                                                          & 100.00\%                                                                                            & 100.00\%                                                                                           \\ \hline
\multicolumn{6}{c}{\textbf{Whitebox-ADVJOINT}}                                                                                                                                                                                                                                                                                                                                                                                                                                                        \\ \hline
\multicolumn{1}{l|}{\textbf{$asv_s/lcnnHalf_s$}} & 1.89\%                           & 99.99\%                                                                                            & 100.00\%                                                                                          & 100.00\%                                                                                            & 100.00\%                                                                                           \\
\multicolumn{1}{l|}{\textbf{$asv_s/lcnnFull_s$}} & 1.18\%                           & 100.00\%                                                                                           & 100.00\%                                                                                          & 100.00\%                                                                                            & 100.00\%                                                                                           \\
\multicolumn{1}{l|}{\textbf{$asv_s/SENet_s$}}    & 9.06\%                           & 100.00\%                                                                                           & 100.00\%                                                                                          & 100.00\%                                                                                            & 100.00\%                                                                                           \\ \hline
\end{tabular}}
\caption{Success rate of our attacks for different values of $\epsilon$ in the white-box setting}
\label{tab:main_exp_white_box}
\vspace{-4mm}
\end{table}

\textbf{Results.} \textit{1) CMSPEC.} We start by presenting the results obtained by conventional feature-domain adversarial approaches against CMs~\cite{liu2019adversarial,zhang2020black} 
to corroborate our claims concerning the inefficacy of such methods (see \S\ref{subsec:existing-attacks}). The results are in Tables~\ref{tab:main_cmspec_wb} \&~\ref{tab:main_cmspec}. In  each row, the number in bold corresponds to the highest success rate for that specific configuration. $\epsilon$ represents the perturbation budget (i.e., the limit on the adversarial noise's $L_\infty$ norm). The columns where $\epsilon$ equals 0 indicate the success rates of spoofed samples that are not adversarially modified. 
Table~\ref{tab:main_cmspec_wb} lists the results for the white-box setting and contains two rows for each shadow CM; the first row ("direct") is the success rate of adversarial features when directly fed into the classifier (the restoration operation is avoided). The second row ("reconstructed") is the success rate when a restoration operation is added.

We vary $\epsilon$ from 0 to 20. For every CM type, the success rate in the white-box setting where the adversarial spectrograms are directly fed into the target system approaches 100\% ($\epsilon = 10$). These scores drop significantly when reconstructing audio from the adversarial spectrograms (``reconstructed'' at $\epsilon = 10$). These discrepancies stem from the effect of the inverse transformations that drastically degrade the quality of the modified signals. For $lcnnFull_s$, the samples still achieve 100\% success rate after reconstruction. We attribute that to over-fitting of the adversarial examples to the CM shadow, which becomes more dominant as $\epsilon$ grows. For such large values of $\epsilon$, the adversarial is significantly different from the original and the change is well-adapted to that CM's idiosyncrasies. Though of a magnitude large enough to withstand the inverse transformation, the adversarials are unable to transfer to other models, as shown in Table~\ref{tab:main_cmspec}. 

The black-box results in Table~\ref{tab:main_cmspec} represent the success rate of the adversarial waveforms against a target CM after restoring them from the spectrograms perturbed in the feature domain using a shadow CM. This is the only experiment where the black-box system includes only a CM component. Since we identify ASVs as an additional factor that further degrades the performance of adversarial attacks on CMs, we exclude them here, so that the results can be used to demonstrate the flaws in the spectrogram-level adversarial generation strategy. 

Compared to the white-box setting, we can see that all shadow architectures barely manage to increase the success rates in the black-box setting compared to the baseline ($\epsilon=0$), where no adversarial perturbation is applied. Hence, the generated adversarial examples are incapable of transferring since their "magnitude" is decreased as a byproduct of reconstruction, validating the claims made in \S\ref{subsec:existing-attacks}. This result is further substantiated by the experiments provided below, showing how ADVCM does manage to achieve concerning success rates. 

\textit{2) ADVCM, ADVSR, \& ADVJOINT.} Tables \ref{tab:main_exp} \& \ref{tab:main_exp_white_box} present the success rates of our novel techniques for generating the adversarials directly in the time domain for the black-box and white-box settings, respectively. We generate adversarial examples based on the three methods discussed in \S\ref{subsec:our_attack}: ADVCM, ADVSR, and ADVJOINT. Recall that ADVSR and ADVJOINT, which extends ADVSR, both require an additional bonafide reference sample belonging to the victim during the optimization phase. We obtain these bonafide samples from the evaluation section of the dataset.

\textit{\underline{Baseline:}} The vulnerability of ASVs to spoofing attacks has been well established. 
To validate that this is the case for our ASVs and spoofed samples as well, we evaluate the success rate of our spoofed (non-perturbed) recordings against the target ASV alone (without a CM), namely $asv_t$. The spoofed examples achieve a high success rate of $71.65\%$. 

In Table~\ref{tab:main_exp}, the target architecture for each row is a combined ASV-CM deployment. We are interested in the baseline case ($\epsilon = 0$), showing the portion of spoofed speech lacking adversarial perturbations that manages to fool such combined deployments. The shown success rates are low (3.64$\%$ or less), which is significantly lower than the success rate against $asv_t$ alone, and explain the false sense of security associated with modern voice authentication systems due to CMs. Industry deployments achieve even much lower success rates~\cite{todisco2019asvspoof}.

\textit{\underline{Our Attacks:}} For adversarially modified samples ($\epsilon > 0$), compared to manipulation in the feature space (see Tables~\ref{tab:main_cmspec_wb} \& ~\ref{tab:main_cmspec}), smaller values of $\epsilon$ are required due to the significantly smaller input audio magnitudes compared to spectrogram values. At $\epsilon = 0.003$ all our approaches yield 100\% white-box success rates (see Table~\ref{tab:main_exp_white_box}).

We start by investigating ADVCM. Compared to the baseline ($\epsilon = 0$), ADVCM significantly degrades the target ASV-CM's defending ability. ADVCM perfectly manages to construct adversarial examples that defeat the shadow CM in the white-box setting ($\epsilon = 0.003$ in Table~\ref{tab:main_exp_white_box}). The strategy faces challenges in the black-box setting (see Table~\ref{tab:main_exp}). The best success rate achieved via ADVCM in the black-box setting against $asv_t/lcnnFull_t$ is $22.73\%$ ($\epsilon=0.001$ and shadow is $SENet_s$). Similarly, for  $asv_t/lcnnHalf_t$, the best success rate is $52.24\%$ ($\epsilon=0.005$ and shadow is $SENet_s$), and for $asv_t/SENet_t$ the best score is $66.58\%$ ($\epsilon=0.003$ and shadow is $lcnnHalf_s$). These numbers are drastically larger than those achieved by non-adversarial spoofing attacks ($\epsilon = 0$), demonstrating the  vulnerabilities of widely used systems and further establishing the soundness of generating the adversarials directly in the time domain. Nonetheless, they are nowhere near the white-box success rates (100\%). This is the result of not taking into account the existence of ASVs. The adversarial noise not only does not aspire to embed the victim user's voiceprint in the audio sample, but can also introduce perturbations that may lead to this signature being removed. 

When shifting our focus to ADVSR, we notice that for $asv_t/lcnnFull_t$ it achieves a maximum success rate of $11.59\%$ ($\epsilon = 0.001$ and shadow is $asv_s$). Its optimal success rate against $asv_t/lcnnHalf_t$ is $36.1\%$ ($\epsilon=0.005$ and shadow is $asv_s$) and against $asv_t/SENet_t$ is $93.57\%$ ($\epsilon=0.007$ and shadow is $asv_s$). These results substantiate the efficacy of ADVSR as a loss function. For $asv_t/SENet_t$, ASVDR achieves the optimal success rate among the three approaches. In this last scenario, this method is able to fool both the target ASV and CM with near perfect success and outperforms both ADVCM and ADVJOINT despite ADVSR's loss function considering only the ASV component and ignoring the CM. In the other two cases (i.e., $asv_t/lcnnHalf_t$ and $asv_t/lcnnFull_t$), ASVDR outperforms ADVCM. 
ADVSR manages to preserve the victim's voiceprint and to eliminate the CM overfitting phenomenon (see~\S\ref{subsec:our_attack}) of ADVCM. Additionally, the loss minimized through this method makes the artificial samples sound more like the human speaker with whom they are associated and, therefore, eliminates machine artefacts, resulting in more natural-sounding universal adversarials that bypass CMs (see~\S\ref{subsec:our_attack}). The significantly increased ability to fool CMs using only ADVSR (compared to $\epsilon = 0$) validates this claim and establishes the efficacy of this method. However, the limitations outlined in~\S\ref{sec:our_attack:joint_loss_function} lead to this method not fulfilling the full potential of adversarial examples. For instance, ADVCM still achieves higher success rates in some cases. 

Finally, we examine our novel ADVJOINT method. For $asv_t/lcnnFull_t$ and $asv_t/lcnnHalf_t$, ADVJOINT achieves the highest success rates among the three methods--- $34.09\%$ ($\epsilon=0.001$ and shadow is $asv_s/SENet_s$) and $73.5\%$ ($\epsilon=0.007$ and shadow is $asv_s/SENet_s$), respectively. The highest success rate achievable by ADVJOINT for $asv_t/SENet_t$ is $91.49\%$ ($\epsilon=0.007$ and shadow is $asv_s/lcnnFull_s$). Note that ADVSR performs better than ADVJOINT in this single instance. We attribute this to the samples crafted with ADVJOINT slightly overfitting to the $lcnnFull_s$ architecture of the CM classifier used to craft them, leading to them exhibiting a lower transferability rate. However, the difference in the performance of the two methods for this target configuration is insignificant. All in all, the ADVJOINT method proves to be able to generate high-quality adversarial examples to bypass real-world deployments. On average, this method significantly outperforms ADVCM and ADVSR.

\textit{\underline{Summary:}} The results clearly expose the vulnerability of modern voice authentication platforms to adversarial samples. Although robust to non-proactive spoofing attacks (attack success rates do not exceed $3.64\%$ for $\epsilon = 0$), these systems can be evaded using our novel attacks that craft adversarials in the input domain. Our first method, ADVCM, which crafts adversarials against shadow CMs only, significantly degrades the performance of these defenses. That said, ADVCM is not the optimal strategy in any scenario. Our second method, ADVSR, takes into account the ASV component while crafting the adversarials to preserve the victim's voiceprint and eliminate CM overfitting. It results in high success rates against joint ASV-CM deployments. Nonetheless, since some telltales of spoofed speech are invisible to ASVs, ADVJOINT achieves the optimal success rates (up to 2x compared to ADVCM and 3x compared to ADVSR) by combining the two approaches. All in all, our findings are concerning, demonstrating how systems guarding valuable resources can be made useless (success rates of up to 93.57\%) through adversarial attacks. 

\subsection{Content Preservation}
\label{subsec:context_pres}
In this section, we evaluate the robustness of our attack w.r.t the transcription unit. Our goal is to demonstrate that our attacks are capable of preserving the textual content of given (spoofed) audio clips, so that they are be transcribed precisely to a required authentication text.

\textbf{Experiment Design:} Note that we rely on optimizing machine-generated (modified) speech to produce high quality spoofed audio. Our attacks assume the existence of state-of-the-art spoofing algorithms, such as Shen et al.'s~\cite{shen2018natural} that are accessible to the attacker. Such algorithms are known for their ability to produce high-quality spoofed speech. Evaluating these algorithms is out of the scope of this paper and we assume the algorithms available to the attacker are of that quality and are able to generate speech that matches given texts as determined by speech-to-text transcribers. Unfortunately, this is not necessarily the case for all the algorithms that were used for creating the spoofed samples in the ASVspoof2019 dataset; while some of the algorithms generate high-quality speech, some are less accurate.

We use the Google speech recognition module, available through the $speech-recognition$ Python package. We choose at random 25 adversarial samples from our previous experiment generated using the ADVJOINT method with $\epsilon = 0.003$ and $asv_s/SENet_s$ as the shadow architecture. The adversarial examples where chosen randomly from those that managed to bypass the target models in the black-box setting, in order to evaluate the success of an end-to-end attack. To neutralize the effect of the quality of the spoofing algorithms on the results and exclude any inaccuracies due to our inability to correctly understand the content in the samples, we choose the following strategy: For each of the chosen adversarial samples, we locate the original sample (without our perturbations) in the dataset. The two samples are passed to the speech recognition module. We test whether the system transcribes both samples identically. By doing so, we guarantee that our adversarial perturbations preserve the textual correctness (and given that the spoofing algorithm generates audio that is correctly transcribed, the attacker's goal is achieved). 

\textbf{Results.} Out of the 25 test pairs, 18 were transcribed to identical contents (72\% success rate). We manually examined the 7 failing pairs. The content in 5 of these 7 is not understandable, neither in the original spoofed sample nor in the corresponding adversarial one. Hence, it is only natural that these samples would fail. Our adversarial engine attempts to remove machine artefacts that are predominant in these samples (they do not sound human-like), which we suspect is the source of the difference in the transcribed contents. The original sample in the sixth pair was transcribed into \textit{"the conference of the first test"}, while after listening to it one clearly realizes this output is incorrect, as the speaker is saying \textit{"the conference is the first test"}. This matches the output given for our adversarial sample. By removing some artefacts, the content became clearer to the speech recognition module. In the seventh pair our adversarial sample indeed fails as it is transcribed to \textit{"no irony was not lost"}, as opposed to the original sample, which is transcribed as \textit{"the irony was not lost"}, matching the sample's actual content. Overall, when discarding the five "inferior" test pairs, our adversarial engine's success rate is 95\% (19/20). 

We further manually verified that in the vast majority of the cases where the pairs were transcribed into the same text, the text indeed matched the content of the audio samples in these pairs. In instances where there was not a match, the differences were minor (a single word mistranscribed). Imperfections in speech-to-text algorithms have warranted the use of an error threshold to allow for slight differences between the requested and the transcribed phrases when performing voice authentication, as done in Voicefox~\cite{voicefox}. We note that the same paper argues that spoofed speech is much more likely to be mistranscribed than bonafide speech, and therefore proposes to use this phenomenon as a defense strategy. However, we did not encounter this issue in our experiments. Regardless, we consider this aspect outside the scope of this work and suggest a possible workaround based on adversarial optimizations in~\S\ref{subsec:mitigation}. 

\subsection{Indistinguishability to a Human Judge}
Our threat model mandates that our attacks must fool human listeners (HJ). Given an audio sample, a claimed identity, and some text, this person is responsible for verifying that the sample 1) comes from the claimed user and 2) matches the textual content. Our attacks optimize synthetic examples generated by state-of-the-art spoofing algorithms. Thus, we avoid evaluating the spoofing algorithms themselves but rather our adversarial attacks. 
We resort to the ABX Listening Test \cite{munson1950standardizing}, where participants are initially required to listen to two samples, $A$ and $B$. Afterward, one of the two samples, $X$, is randomly chosen and replayed to the participant, who has to determine whether $X$ is $A$ or $B$. 

In our study, $A$ and $B$ are a spoofed sample and an adversarial recording generated from it (or vice versa). The goal is to show  that a listener cannot identify the replayed sample with a better than random success probability. This setup implies far more restrictive constraints on our adversarial samples than needed for our attack: we need the adversarial samples to only sound as if they were uttered by the victim and preserve the content. Yet, we here demand that they remain \textbf{identical} to the original samples. 

We ran our IRB-approved user study on MTurk~\cite{amazon} using the same 25 test pairs from \S\ref{subsec:context_pres}. Following ABX guidelines~\cite{abx}, we limited the study to 25 samples to avoid faulty responses due to fatigue or lack of concentration. Additionally, to ensure high-quality responses, we filtered out participants with MTurk approval rates below 95\%. We received 20 responses, out of which 19 contained answers for the 25 ABX tasks. One respondent chose to quit after completing 20 questions. To ensure the results for each pair are not influenced by its position in the test (toward the end participants become less focused), we shuffled the tasks for each subject.

For the 25 test pairs, we received a total of 495 responses, which were pooled per sample pair and converted to a proportion of correct responses out of the total responses for each pair. To verify the samples are not distinguishable to human listeners, we formulate a hypothesis test under the assumption that the responses were selected randomly. We test our null hypothesis $H_0: \mu = 0.5$, where $\mu$ is the true probability of selecting the correct response for each test pair, using a one sample t-test. We fail to reject the null hypothesis with a significance level of $\alpha=0.05$ and a p-value of 0.142. The mean proportion for the 25 pairs was 0.54 with a 95\% confidence interval [0.486, 0.594]. We conclude that there is not sufficient evidence against the assumption that the responses were randomly selected, establishing the efficacy of our attacks and their ability to fool human listeners.


\section{Over-Telephony-Network Attacks}
\label{sec:Over-Telephony}
We now present our adversarial attacks on CMs for the over-telephony-network scenario. To the best of our knowledge, we are the first to explore this setting for targeted attacks---the only known adversarial attack against an audio system over a telephony network is non-targeted (speech is perturbed so that it is arbitrarily misclassified)~\cite{overtelephony}. We present superior targeted attacks with the purpose of impersonating a specific victim  to both machines and humans. Attacks over the telephony network have been proven to be harder to mount due to three main challenges: codecs, packet loss and jitter~\cite{abdullah2020faults}. 

A codec is used to encode and decode audio signals before/after transmitting them over the network. 
Modern VOIP and VOLTE service providers employ codecs that compress the data before transmitting it. Codecs used in practice are lossy. 
When choosing adversarial examples as the attack strategy, this property significantly raises the bar, as the adversarial perturbations are volatile and, thus, are not guaranteed to survive the lossy (noisy) encoding/decoding procedure. 

The second factor affecting the success rates of our adversarial examples is the network's packet loss, which leads to the received data not perfectly matching the transmitted signal even in the presence of a lossless codec. A loss of up to 2\% of packets is normal. Adversarial examples over telephony networks should be robust against packet loss.

Finally, there is the jitter effect. Modern codecs are equipped with jitter buffers that minimize the effects of jitter. Therefore, jitter is not a factor considered in our experiments.

\subsection{Simulating Over-Telephony-Network Attacks}
To simulate realistic attacks, we choose the open-source Opus codec \cite{opus}. Opus is a high-quality lossy codec used by WhatsApp. 
Additionally, it is on par with the Enhanced Voice Services (EVS) audio coding standard~\cite{evs} that is supported by numerous phone carriers and devices. Hence, our findings are applicable to phone as well as VOIP calls. 

To mount our attacks, we generate the adversarials, encode them with Opus, transfer them from a remote datacenter over the Internet to a server at our university using UDP, decode them and feed them into the target system. 
The Opus repository \cite{opus} contains a sub-project, \textit{opus-tools}, which offers scripts that facilitate our first-of-its-kind experiment. First, the script \textit{opusenc} takes an audio waveform and outputs a $.opus$ file corresponding to the encoded signal by the tool. Second, \textit{opusrtp} receives the opus-encoded signal and sends it over $RTP$ (Opus's application layer protocol over UDP) to the destination. At the destination, we can use $opusdec$, which takes as input the received $opus$-encoded file extracted from $pcap$ packets and decodes it back to a waveform. The tool also allows us to set the percentage for the packet loss we wish to simulate at this stage. All our experiments specify a 2\% packet loss (to simulate real-world networks). Sending the samples between two servers far away from each other introduces additional packet loss and jitter.

\subsection{Adversarial Example Generation}
 As explained above, transmitting adversarial examples over the telephony network is more challenging due to packet loss and codecs. To overcome the packet loss, we add redundancy to the encoded packets---Opus offers the option 
to automatically generate redundancy to enable recovering from an expected percentage of packet loss (2\% in our experiments). There remains the use of lossy codecs. This loss is typically the main challenge for adversarial attacks, and it occurs due to the codecs' inability to decode the compressed signal with 100\% accuracy. These discrepancies between the original and the reconstructed signal can be viewed as noise that may neutralize the effect of the adversarial perturbations. Our attacks, however, are naturally robust to this form of distortion. The reason is that codecs perform lossy voice compression through psycho-acoustic techniques to discard useless information and obtain a compact representation of the input audio, as determined by humans~\cite{herre2019psychoacoustic}. Our attacks optimize audio samples to deceive CMs by inserting human-like "fingerprints" or eliminating machine artefacts. Thus, we expect our adversarial modifications to be highly compatible with audio codecs, which will deem them necessary (to \textbf{humans}), leading to them not being discarded. Since our time-domain adversarials are effective in the over-the-line setting, they remain so when transmitted over the network. That said, some loss is to be anticipated. To overcome this, we propose the following strategy: We let the function $\eta : X \longrightarrow X$ model the encoding/decoding distortion. To generate an adversarial from utterance $x$, we first pass it through the encoding/decoding stages to obtain the noisy $\eta(x)$, which resembles the signal to be received at the remote system and encapsulates the added decoder noise. At this point, $\eta(x)$ is fed into the shadow CM and the adversarial noise $\delta$ is crafted w.r.t $\eta(x)$. Finally, $x + \delta$ is forwarded to the server. Since $\delta$ represents an imperceptible minor perturbation, we expect $x$ and $x + \delta$ to experience the same noise when propagated through the codec, or mathematically
\begin{equation*}
    \lim_{\delta \longrightarrow 0} \eta(x + \delta) = \eta(x) + \delta.
\end{equation*}

Since we craft $\delta$ such that $\eta(x) + \delta$ is a valid adversarial example, we expect $\eta(x + \delta)$ that is received at the server (since we encode $x + \delta$ to which the decoding noise is added) to also fool the target system.

\subsection{Experiments and Results}
Table~\ref{tab:codecs} shows the black-box results of our over-telephony-network experiments, which involve the same 20K pairs of samples from~\S\ref{subsec:fooling_the_machine}. We generate attacks with ADVJOINT, since we proved it to be superior to the other methods. 
In these experiments, we switch the target and shadow models used in the earlier experiments to demonstrate that our results do not manifest themselves only for the initial configuration. These experiments also include a state-of-the-art GMM x-vector model as the target ASV, to demonstrate the ability of our adversarials to transfer to these architectures as well. The target systems, hence, are ASV-CMs, where the ASV may be either a GMM i-vector (which is identical to $asv_s$ from earlier), or a GMM x-vector, trained on a different dataset (the VoxCeleb dataset~\cite{nagrani2017voxceleb}). The initial EER values for this model on the test partition of VoxCeleb are in Table~\ref{table:eer} in Appendix~\ref{app:models}. Our GMM x-vector architecture is taken from~\cite{li2020adversarial}.

For each value of $\epsilon$ we have two columns (i-vector/x-vector), and the rates in these columns refer to the success probability against the target ASV-CM configuration wherein the ASV model is either the i-vector or the x-vector. The CM target is in the second column. The success rates of the non-adversarial spoofing attacks ($\epsilon=0$) demonstrate that the target ASV-CMs are robust. Interestingly, we find that x-vector ASVs are more vulnerable to spoofing attacks than their i-vector counterparts. The reason behind this is that the x-vector model is a DNN with more parameters and that it is trained on more data, making it more accurate. Since spoofing attacks are meant to sound like the victim, the more "accurate" x-vector performs "better" by being more vulnerable to spoofing attacks (it accepts the fake samples that sound like the victim). Specifically, Li et al.~\cite{li2020adversarial} show how the x-vector model outperforms the i-vector one when both are trained on VoxCeleb. 
Nonetheless, when increasing $\epsilon$, both systems become highly vulnerable. The highest success rates are obtained at the largest value of $\epsilon$ (0.007). Remember that in \S\ref{subsec:fooling_the_machine} the highest success rates for most target systems were measured at lower values of $\epsilon$ (0.001). The reason is that in this experiment the attacks are transmitted over the network and the adversarial perturbations are "removed" due to various factors mentioned earlier. To be able to still affect the target models' decision, there is a need for a larger perturbation magnitude. Increasing the value of $\epsilon$ beyond 0.007 may result in better success rates. However, this will diminish the quality of the spoofed speech. Finding a sweet spot is left to future work.

Our experiments demonstrate the practicality of over-telephony-network adversarial attacks, achieving success rates as high as 37.2\% (compared to a baseline of 5.79\%), and call into question the reliability of deployed voice authentication systems.  

\begin{table*}
\centering
\resizebox{\textwidth}{!}
{\begin{tabular}{@{}clllllllllllll@{}}
\toprule
                             &                     & \multicolumn{2}{c}{\textbf{$\epsilon$ = 0}} & 
                                                   \multicolumn{2}{c}{\textbf{$\epsilon$ = 0.001}} &   
                                                   \multicolumn{2}{c}{\textbf{$\epsilon$ = 0.003}} &   \multicolumn{2}{c}{\textbf{$\epsilon$ = 0.005}} &
                                                   \multicolumn{2}{c}{\textbf{$\epsilon$ = 0.007}}   \\ \midrule

\multicolumn{1}{c}{\textit{Shadow}} & {\backslashbox{\textit{Target CM}}{\textit{Target ASV}}}    

& \multicolumn{1}{c}{\textbf{i-vector }} &
\multicolumn{1}{c}{\textbf{x-vector}}  & \multicolumn{1}{c}{\textbf{i-vector }} &
\multicolumn{1}{c}{\textbf{x-vector}}  & \multicolumn{1}{c}{\textbf{i-vector }} &\multicolumn{1}{c}{\textbf{x-vector}}  & 
\multicolumn{1}{c}{\textbf{i-vector }} & \multicolumn{1}{c}{\textbf{x-vector}}  &
\multicolumn{1}{c}{\textbf{i-vector }} & \multicolumn{1}{c}{\textbf{x-vector}}  &   \\ \midrule
\multicolumn{1}{c|}{\textbf{$asv_t/lcnnHalf_t$}} & \multicolumn{1}{c|}{\textbf{$lcnnFull_s$}} & 3.53\%                    & \multicolumn{1}{l|}{5.79\%} & 7.3\%                    & \multicolumn{1}{l|}{10.015\%}       & 15.3\%                                & \multicolumn{1}{l|}{18.705\%}          & 24.19\%                             & \multicolumn{1}{l|}{28.45\%}           & \textbf{32.22}\%                                  & \textbf{37.205}\%             \\
\multicolumn{1}{c|}{\textbf{$asv_t/lcnnHalf_t$}}  & \multicolumn{1}{c|}{\textbf{$SENet_s$}} & 1.82\%                    & \multicolumn{1}{l|}{5.345\%}     & 1.97\%                             & \multicolumn{1}{l|}{3.035\%}       & 5.935\%                             & \multicolumn{1}{l|}{7.37\%}          &  9.065\%                             & \multicolumn{1}{l|}{10.825\%}       &  \textbf{9.965\%}                        & \textbf{11.555}\%             \\
\multicolumn{1}{c|}{\textbf{$asv_t/lcnnFull_t$}} & \multicolumn{1}{c|}{\textbf{$lcnnHalf_s$}} & 3.04\%                    & \multicolumn{1}{l|}{8.4\%}     & 6\%                             & \multicolumn{1}{l|}{10.935\%}   
& 11.275\%                             & \multicolumn{1}{l|}{15.655\%}          & 14.72\%                    & \multicolumn{1}{l|}{18.68\%}       &  \textbf{18.355}\%                                 & \textbf{22.03}\%             \\
\multicolumn{1}{c|}{\textbf{$asv_t/lcnnFull_t$}} & \multicolumn{1}{c|}{\textbf{$SENet_s$}} & 1.82\%                    & \multicolumn{1}{l|}{5.345\%}      & 3.355\%                             & \multicolumn{1}{l|}{5.3\%}       & 11.005\%                             & \multicolumn{1}{l|}{13.745\%}         & 15.32\%                             & \multicolumn{1}{l|}{18.225\%}       &  \textbf{16.745\%}                        & \textbf{19.205}\%             \\
\multicolumn{1}{c|}{\textbf{$asv_t/SENet_t$}} & \multicolumn{1}{c|}{\textbf{$lcnnHalf_s$}}   & 3.04\%                    & \multicolumn{1}{l|}{8.4\%}    & 6.19\%                             & \multicolumn{1}{l|}{10.095\%}       & 11.77\%                             & \multicolumn{1}{l|}{15.075\%}          & 16.935\%                             & \multicolumn{1}{l|}{20.255\%}        & \textbf{21.38}\%                                 & \textbf{24.99\%}    \\
\multicolumn{1}{c|}{\textbf{$asv_t/SENet_t$}} & \multicolumn{1}{c|}{\textbf{$lcnnFull_s$}}   & 3.53\%                    & \multicolumn{1}{l|}{5.79\%}      & 7.74\%                             & \multicolumn{1}{l|}{9.87\%}       & 14.37\%                                & \multicolumn{1}{l|}{16.38\%} &  21.855\%                             & \multicolumn{1}{l|}{24.305\%}       & \textbf{28.08}\%                                  & \textbf{31.035}\%             \\ \bottomrule
\end{tabular}}
\caption{Success rates of over-telephony-network attacks for different values of $\epsilon$ in the black-box setting (ADVJOINT)}
\label{tab:codecs}
\vspace{-4mm}
\end{table*}

\section{Discussion}
\label{sec:Discussion}
\subsection{Practicality of Our Attacks}
\label{sec:discussion:practicality}
Generating the adversarial perturbations in all our attacks takes on average 4 seconds per sample, making them highly practical, as the attacker can mount them in real time and respond to an authentication request within an amount of time consistent with that expected from a human. The attacker also needs to generate the synthetic speech beforehand, which takes milliseconds with state-of-the-art VC and SS algorithms~\cite{nuancetts}. 

\subsection{Mitigation of Our Attacks}
\label{subsec:mitigation}
Liveness detection techniques that rely on physical features that accompany human speech, such as VoiceLive~\cite{zhang2016voicelive}, VoiceGesture~\cite{zhang2017hearing} and EchoVib~\cite{echovib}, are mitigation strategies against our attacks since they detect the absence of a human speaker. Nonetheless, these methods may not always be applicable as they are device-dependent and require the support of the device used to interact with the service. For instance, in the over-telephony-network scenario, the back-end servers need to support users who call from landlines as well, which are not equipped with sensors required for these defenses. An attacker with access to the device's internals may also be able to mount powerful attacks to circumvent these methods. That said, the fusion of these defenses may be a promising approach.

Another possible class of defenses involves adversarial attacks' mitigation techniques, and specifically spatial smoothing and adversarial training~\cite{wu2020defense}, self-supervised learned adversarial defenders~\cite{wudefblack}, and audio squeezing. Nonetheless, as FAKEBOB~\cite{chen2019real} shows, spatial smoothing and audio squeezing are ineffective against attacks on acoustic systems. Wu et al.~\cite{wudefblack} assume adversarials generated in the feature space and likely have limited applicability against our attacks. Adversarial training and similar approaches~\cite{pal2020adversarial, joshi2021adversarial, wang2019adversarial, gan, kingma2013auto} are typically incapable of protecting against attacks derived with different adversarial algorithms and parameters~\cite{li2020investigating}.

Finally, Voicefox~\cite{voicefox} suggests that transcribers may be used to spot machine generated audio to defeat speech synthesis attacks. However, the attacker may adversarially attempt to optimize the generated speech to fool speech-to-text components as well. Exploring this phenomenon is left to future work.

Given the vulnerability of voice authentication to adversarial examples, we contend it must be accompanied by another authentication factor, such as facial recognition or fingerprints.

\subsection{Future Work}
Our attacks require access to some rooted device to inject the adversarial waveforms. 
This may be redundant if we prove the attacks' ability to survive in the over-the-air setting. Earlier works have demonstrated that audio adversarial samples may survive transmission through a microphone~\cite{chen2019real}. In this case, the attacker may even use the victim's device for the attack. In the over-telephony-network setting, our attacks must withstand the network distortion. Quantifying the degradation as a result of transmission through two lossy mediums sequentially and developing methods to maintain high success rates is a future research avenue. Developing verification-oriented ML classifiers for authentication that are by nature robust to adversarial examples is another direction we are currently exploring.

\section{Related Work}
\label{sec:RelatedWork}

The literature commonly discusses spoofing attacks in two main settings: Logical Access (over-the-line) and Physical Access (over-the-air)~\cite{wu2017asvspoof}, and countermeasures have been proposed for both. In the over-the-air setting, that is, when spoofed samples have to be played through speakers to reach the target ASV, various methods targeting physical differences between audio produced by the human vocal system and speakers have been employed as telltales for detecting spoofing attacks. These include articulatory gestures of human users~\cite{zhang2017hearing},  time-difference-of-arrival (TDoA) changes in a sequence of phoneme sounds to the two microphones of the phone~\cite{zhang2016voicelive}, artefacts unique to spectrograms of audio played through speakers~\cite{ahmed2020void}, and vibrational signatures~\cite{echovib}.

We consider the more problematic scenario for defenses, where the modified audio need not be played through speakers. This is referred to as the logical access setting. Here, the threat model limits tools at our disposal for spotting spoofed speech to methods that do not rely on the physical properties of electronic speakers. The ASVspoof challenge~\cite{wu2017asvspoof} has long been the platform at which such novel CM architectures emerge, targeting both the logical and physical access settings. However, existing studies have proven CMs to be vulnerable to certain attacks themselves~\cite{wu2017asvspoof, kinnunen2018spoofing}. 
Yet, more advanced CMs have since appeared, demonstrating increased robustness~\cite{wu2017asvspoof}. The above are naive, non-proactive attacks that do not target ASVs or CMs, but rather circumvent these systems as an (un)fortunate byproduct of a benign technology. 

More powerful, task-oriented attacks later emerged with the introduction of adversarial examples. Adversarial attacks on acoustic systems~\cite{carlini2018audio, alzantot2018did, cisse2017houdini} have been known since 2017. The first work to explore adversarial examples on ASVs was presented by Kreuk et al. \cite{kreuk2018fooling}, where the authors introduced perturbations to the spectral representations of speech signals. Afterward, the inverse transform was applied to recover the adversarial audio. The same methodology was later adopted by Li et al.~\cite{li2020adversarial}, who also considered GMM i-vector/x-vector models and investigated the transferability of adversarial speech across ASV architectures. FAKEBOB~\cite{chen2019real} demonstrated attacks in a fully black-box scenario, using a threshold estimation procedure. Recent works~\cite{xie2020real, li2020practical} explore the practicality of over-the-air attacks by modeling the Room Impulse Response, but are limited to the white-box setting. Jati et al.~\cite{jati2021adversarial} compared various adversarial attacks and evaluated the efficacy of known defenses. Other works \cite{villalba2020x, wang2020inaudible} focus mainly on x-vector-based models and their vulnerability to adversarial examples. SirenAttack~\cite{du2020sirenattack} operates in the logical access setting and introduces a method to craft adversarials against automatic speaker recognition systems assuming the waveforms can be directly fed into the target. This assumption has also been made in other previous works~\cite{carlini2018audio, abdullah2019practical}. However, spoofing countermeasures are not considered in~\cite{du2020sirenattack}. AdvPulse~\cite{li2020advpulse} is concerned with the physical access scenario and proposes a penalty-based approach combined with noise modeling to generate universal over-the-air adversarials. For a broader discussion of adversarial attacks on ASVs, we refer to Abdullah et al.~\cite{abdullah2020faults}. These attacks all fail to generate samples that sound like the victim and fail under our threat model (\S\ref{sec:Threat}).

Adversarial attacks on CMs have so far received less attention. The first work is due to Liu et al.~\cite{liu2019adversarial}, where spectral features are manipulated by shadow models to craft adversarial examples. The authors consider a semi-black-box setting through transferability across different architectures. Zhang et al.~\cite{zhang2020black} aimed at enhancing the transferability using an iterative optimization and an ensemble of shadow models. These works do not include target ASVs. Moreover, the adversarials are generated in the feature space, and therefore are incapable of posing threats in real-world scenarios (\S\ref{subsec:existing-attacks}).  

When discussing CMs and ASVs, it is essential to remember that the two are to be deployed jointly. Nonetheless, attacks targeting a combined ASV-CM system have not emerged yet. 
The literature presents several possible integration strategies for these systems, although the robustness of these methods is evaluated using only non-proactive attacks against which CMs are trained. A cascaded approach was first introduced by Sahidullah et al.~\cite{sahidullah2016integrated}, while Todisco et al.~\cite{todisco2018integrated} later proposed a more advanced Gaussian back-end fusion scheme to combine both models, and Dhanush et al.~\cite{dhanush2017factor} presented a joint approach for learning the tasks simultaneously using factor analysis methods. 
Gomez-Alanis et al.~\cite{gomez2020joint} suggested an integration neural network. Aljasem et al.~\cite{sm} proposed an integrated defense via a novel set of features and ensemble learning through asymmetric bagging. None of these works introduce attacks specifically targeting the integrated system. 

\section{Conclusion}
We presented the first practical adversarial attack on spoofing countermeasures. Our novel algorithms are capable of generating high-quality spoofed speech that fools both machines and humans, and enables the attacker to pose as the victim in security-critical environments and severely compromise users' security.
The success rates of our attacks are concerning, primarily due to them being attained in the black-box setting and under the assumptions of realistic threat models, including the smartphone app scenario and the phone conversation setting, for which we also demonstrated novel targeted attacks. Our findings highlight the severe pitfalls of voice authentication systems and stress the need for more reliable mechanisms.

\section*{Acknowledgment}   
We gratefully acknowledge the support of the Waterloo-Huawei Joint Innovation Laboratory for funding this research. We would also like to thank the Compute Canada Foundation (CCF) for their resources that made our experiments possible.

\bibliographystyle{plain}
\bibliography{Sections/refs}{}

\appendix
\section{Appendix}
\subsection{Baseline EER Values for The CM and ASV Models}
\label{app:models}
Table~\ref{table:eer} provides the individual EER values for the various CMs and ASVs with which we experiment.

\mathchardef\mhyphen="2D

\begin{table}[h]
\centering
\begin{tabular}{llll}
\hline
\multicolumn{2}{c|}{\textbf{CM}}                 & \multicolumn{2}{c}{\textbf{ASV}} \\ \hline
\multicolumn{4}{c}{\textbf{Target Models}}                                          \\ \hline
\textbf{$lcnnHalf_t$} & \multicolumn{1}{l|}{8.06\%}  & \textbf{$asv_t$}      & 2.76\%             \\
\textbf{$lcnnFull_t$} & \multicolumn{1}{l|}{4.36\%}  & \textbf{$x \mhyphen vector_t$}      & 6.24\%                    \\ 
\textbf{$SENet_t$}    & \multicolumn{1}{l|}{7.73\%}  &             \\ \hline
\multicolumn{4}{c}{\textbf{Shadow Models}}                                          \\ \hline
\textbf{$lcnnHalf_s$} & \multicolumn{1}{l|}{6.16\%}  & \textbf{}          &            \\
\textbf{$lcnnFull_s$} & \multicolumn{1}{l|}{5.11\%}  & \textbf{$asv_s$}      & 3.0\%       \\
\textbf{$SENet_s$}    & \multicolumn{1}{l|}{10.92\%} &                    &             \\ \hline
\end{tabular}
\caption{Baseline CM and ASV target and shadow model EER values (w.r.t the corresponding test sets)}
\label{table:eer}
\end{table}

\subsection{Target CM and Attacker using Identical Spoofing Algorithms}
\label{app:trained_on_eval} Unfortunately, there is a shortage in datasets containing synthetic and modified speech generated with a variety of algorithms that are suitable for training state-of-the-art CMs, wherein multiple speakers participate and their labels are retained. Since ASVspoof2019 is the gold standard for such a dataset, we made the choice to utilize the entirety of the dataset in our experiments. This was done by using the training and development sections to train the shadow and target models, respectively, and the evaluation section to craft adversarial samples. We realize that this may raise concerns regarding the practicality of our threat model, as one might think that since the training and development sections contain synthetic speech generated with the same algorithms, the attacker in our scenario may have leverage that a real-world adversary does not. 

This section serves to establish that the threat model we present in~\S\ref{sec:Threat} is realistic and practical. To that end, we demonstrate that the assumption we make in~\S\ref{subsec:attacker_knowledge}, which is that the attacker knows the algorithms used to generate the spoofed speech to train the target CMs, does \textbf{not} result in a more powerful attacker than a real-world adversary. That is, we wish to prove that since state-of-the-art SS and VC algorithms all share similar underlying principles, having access to the same algorithms as the target CM does not result in a stronger attack. To do so, we compare two attackers: the first has access to the same algorithms for generating spoofed speech as the target CM, while the second does not. The second attacker will naturally use the same algorithms for both generating spoofed speech to train the shadow models and for creating fake samples to mount the adversarial attacks. 
By showing that the second attacker can achieve success rates comparable to the first who does know the algorithms used by the target system, we cement the claim that this knowledge is of limited value, if any. Hence, the attacker in our threat model is as strong as a real-world malicious party. 

We present the results of an experiment where the shadow CMs were trained on a subset of the evaluation section of ASVSpoof2019. The spoofed samples in the evaluation section were all generated with spoofing algorithms different from the algorithms used to create the spoofed samples in the training and development sections. The chosen subset contains 25,474 (22,932 spoofed/2,542 bonafide) samples from 23 speakers. The number of speakers and samples is on par with the sizes of the development and training sets we use in the main experiments in \S\ref{subsec:fooling_the_machine}. The models' baseline performance (in the absence of adversarial attacks) is shown in Table~\ref{table:eer_eval}, and was obtained by evaluating the models on all 
the samples from the evaluation section. We refer to these models as the \textit{eval. shadows} and identify them by an $e$ subscripted to their architecture's name. The shadow models from \S\ref{subsec:fooling_the_machine} are denoted as the \textit{train shadows} and are subscripted with an $s$. Their baseline performance is in Table~\ref{table:eer}. Clearly, the \textit{eval. shadows} achieve lower EERs and outperform the \textit{train shadows}. This observation can be easily reasoned about since the \textit{eval. shadows} were trained and evaluated on data generated using the same spoofing algorithms. However, the data used for training the \textit{train shadows} was generated using  spoofing algorithms different from the ones used for generating the data on which the \textit{train shadows} were evaluated.

\begin{table}[h]
\centering
\begin{tabular}{llll}
\hline
\multicolumn{4}{c}{\textbf{CM}}                 \\ \hline
\textbf{$lcnnHalf_e$} & \multicolumn{2}{l}{1.493\%}             \\
\textbf{$lcnnFull_e$} & \multicolumn{2}{l}{1.385\%}  & 
\\
\textbf{$SENet_e$}    & \multicolumn{2}{l}{1.85\%}  &             \\ \hline
\end{tabular}
\caption{EER values for the CM shadow models trained on a subset of the evaluation section of ASVspoof2019 (\textit{eval. shadows})}
\label{table:eer_eval}
\end{table}

Next, we use ADVJOINT, which is the optimal adversarial examples generation strategy we propose. 
The key difference between this experiment and the one in \S\ref{subsec:fooling_the_machine} is that here the shadow CMs are trained on spoofed samples generated using the \textbf{same algorithms that are later used to craft attacks}. On the other hand, in \S\ref{subsec:fooling_the_machine}, the shadow CMs are trained on data generated using the algorithms that were also used to generate the data to train the target models. It is natural to assume that the attacker would be able to train the shadow CMs with the same algorithms later used to mount their attacks. This approach was used in this experiment, but not in the experiment in \S\ref{subsec:fooling_the_machine}. There, the training of the shadow CMs exploited the training section of the ASVspoof2019 dataset, and the generation of the adversarials was based on its evaluation section. As mentioned above, the spoofing algorithms used to create the spoofed samples in the training section were different from the spoofing algorithms used for crafting the spoofed samples in the evaluation section.

Out of the 20K spoofed examples used in \S\ref{subsec:fooling_the_machine}, we keep 12,764 that do not appear in the subset selected for training the \textit{eval. shadows} and use only these samples to generate adversarial attacks. We do so since we wish to model a real-world scenario. Table~\ref{tab:trained_on_eval_attack}  presents the success rates of the attacks (using ADVJOINT) for $\epsilon = 0.001$. We compare the results to those obtained using the same method and the same $\epsilon$, but with the \textit{train shadows}, in Table~\ref{tab:main_exp}. We choose $\epsilon = 0.001$ since it was found to be optimal in three out of the six target configurations in \S\ref{subsec:fooling_the_machine}.

\begin{table}[h]
\centering
\begin{tabular}{llllll}
\hline
\multicolumn{1}{l}{\textbf{}}              &  \multicolumn{1}{c}{\textit{\textbf{\begin{tabular}[c]{@{}c@{}}$\epsilon$\\  =0\end{tabular}}}} & \multicolumn{1}{c}{\textit{\textbf{\begin{tabular}[c]{@{}c@{}}$\epsilon$\\  =0.001\end{tabular}}}} 
\\ \hline
\multicolumn{2}{c}{\textbf{Blackbox}}                                                                                                                                                                                                                                                                                                                                                                                                                                                       \\ \hline
\multicolumn{1}{l|}{\textbf{$asv_s/lcnnHalf_e - asv_t/lcnnFull_t$}}          & 1.37\%                           & 15.1\%                                                                 \\ 
\multicolumn{1}{l|}{\textbf{$asv_s/lcnnHalf_e - asv_t/SENet_t$}}          & 2.984\%                           & 48.32\%                                                                 \\ 
\multicolumn{1}{l|}{\textbf{$asv_s/lcnnFull_e - asv_t/lcnnHalf_t$}}          & 3.49\%                           & 45.41\%                                                                 \\ 
\multicolumn{1}{l|}{\textbf{$asv_s/lcnnFull_e - asv_t/SENet_t$}}          & 2.984\%                           & 70\%                                                                 \\
\multicolumn{1}{l|}{\textbf{$asv_s/SENet_e - asv_t/lcnnFull_t$}}          & 1.37\%                           & 17.6\%                                                                 \\
\multicolumn{1}{l|}{\textbf{$asv_s/SENet_e - asv_t/lcnnHalf_t$}}          & 3.49\%                           & 29\%                                                                 \\
\hline
\end{tabular}
\caption{Success rate with shadows trained on the same algorithms used for crafting attacks}
\label{tab:trained_on_eval_attack}
\end{table}

Note that the unmodified spoofed examples' success rates ($\epsilon = 0$) remain somewhat identical to those in Table~\ref{tab:main_exp}, which is to be expected, since these are the same samples evaluated on the same targets, excluding the ones we discarded since they appear in the training dataset of our \textit{eval. shadows}. For the configuration where the shadow CM is $lcnnHalf$ and the target is $lcnnFull$, the attack's success rate for $lcnnHalf_e$ is 15.1\%, compared to 12.22\% for $lcnnHalf_s$. For the shadow CM $lcnnHalf$ and target $SENet$, $lcnnHalf_e$'s success rate is 48.32\%, compared to 51.32\% by $lcnnHalf_s$. For $lcnnFull$ against $lcnnHalf$, we achieve a success rate of 45.41\% using $lcnnFull_e$, where in the previous experiment the success rate of $lcnnFull_s$ was 37.29\%. For $lcnnFull$ attacking $SENet$, we obtain a success rate of $70\%$ by $lcnnFull_e$, which is a significant improvement over the 41.1\% success rate reached by $lcnnFull_s$. Interestingly, when SENet is employed as the shadow, we notice that training the shadow on the training subset resulted in higher success rates: $SENet_s$ achieves a success rate of 34.09\% against the $lcnnFull$ target, compared to the 17.6\% reached by $SENet_e$, and 46.3\% against $lcnnHalf$ compared to 29\% ($SENet_e$). Nonetheless, when considering all three shadow architectures, it is easy to see that none of the two shadow sets (\textit{eval. shadows} and \textit{train shadows}) have a clear advantage over the other. This is despite the data used to train the \textit{train shadows} coming from the "same data distribution" or, in other words, being crafted with the same algorithms for speech synthesis and conversion as the data used to train the target models in our black-box setting, which is not the case for the \textit{eval. shadows}. 

We conclude that the assumption of the attacker having access to samples generated with the same algorithms as those used by the target CM does not make the attacker's task easier. Having access to different algorithms but using them both for creating training samples and for the generation of the adversarials is as powerful. It is useful to gain access to training samples generated with the same algorithms that are later also used to craft the adversarials since the trained shadow CMs better learn the specific artefacts of the relevant algorithms and better optimize the adversarial samples they yield. This sort of samples is trivially available to a real-world attacker, who can determine the algorithm to use for creating adversarials based on the samples available to them during training.


\end{document}